\documentclass[11pt,a4paper]{article}

\usepackage{jcappub}
\usepackage{bm}
\usepackage{multirow,array}
\usepackage{comment}
\usepackage{slashed}
\newcommand{\dif}{\mathrm{d}}
\newcommand{\del}{\partial}

\newcommand{\e}{\mathrm{e}\,}
\newcommand{\im}{\mathrm{i}\,}

\title{Chiral asteroseismology: seismic oscillations caused by chiral transport in neutron stars and supernovae}
\author{Sota Hanai}
\author{and Naoki Yamamoto}
\affiliation{Department of Physics, Keio University,\\
3-14-1 Hiyoshi, Yokohama, Japan}
\emailAdd{souta.h.0619@keio.jp}
\emailAdd{nyama@rk.phys.keio.ac.jp}
\abstract{
We study the novel asteroseismology of the chiral magnetic wave (CMW) of the quark number density in relativistic quark matter inside neutron stars and core-collapse supernovae and the chiral vortical wave (CVW) of the neutrino number density in relativistic neutrino matter at the core of supernovae. 
We call the oscillation modes for these chiral waves the chiral magnetic mode (CM-mode) and chiral vortical mode (CV-mode), respectively.
We derive the dispersion relations of these new modes in the presence of the chirality flipping due to the finite quark mass and diffusion. We then estimate the possible frequencies of these modes and amplitudes of the resulting gravitational waves. In particular, since the CM-mode can exist only in quark matter with nearly gapless quarks (such as the two-flavor color superconductivity) for a sufficiently strong magnetic field, corresponding gravitational waves provide a new possible probe for such quark matter and the magnetic field in neutron stars.
}
\arxivnumber{2203.16133}
\begin{document}
\maketitle

\section{Introduction}
\label{sec:introduction}

Unraveling the internal structure of neutron stars and core-collapse supernovae is an important problem in astrophysics. One way to estimate the interior of stars is by analyzing seismic oscillations, called asteroseismology. In particular, the possible detection of gravitational waves associated with seismic oscillations in neutron stars and supernovae gives a clue to the interior of these compact stars. In 2015, gravitational waves from binary black holes were detected at the Laser Interferometer Gravitational-Wave Observatory~(LIGO) and Virgo for the first time \cite{LIGOScientific:2016aoc}, and in 2017, gravitational waves from binary neutron stars were also observed \cite{LIGOScientific:2017vwq}. In this stream, gravitational wave asteroseismology is becoming a promising direction for exploring the nature inside neutron stars and supernovae.

In asteroseismology, seismic oscillations and resulting gravitational waves are classified into several oscillation modes according to their physical origins. Typical examples are summarized in section~\ref{sec:asteroseismology}. Because these modes generally depend on the transport properties of stars, understanding the transport phenomena inside neutron stars and supernovae is essential for the classification of possible oscillation modes there.

Recently, new types of transport phenomena due to the chirality of relativistic fermions have attracted considerable interest. The prototype examples are the so-called chiral magnetic effect~(CME) \cite{Vilenkin:1980fu,Nielsen:1983rb,Fukushima:2008xe} and  chiral vortical effect~(CVE) \cite{Vilenkin:1979ui,Son:2009tf,Landsteiner:2011cp}, which are vector currents in the presence of chirality imbalance along a magnetic field and vorticity, respectively; see, e.g., eqs.~(\ref{eq:CME}) and (\ref{eq:CVE}) below for the explicit expressions.
Furthermore, these chiral transport phenomena lead to new collective excitations, such as the chiral magnetic wave (CMW) \cite{Newman:2005hd,Kharzeev:2010gd} and chiral vortical wave (CVW) \cite{Jiang:2015cva}.%
\footnote{There are also other chiral waves, such as the chiral Alfv\'en wave~\cite{Yamamoto:2015ria}, chiral heat wave~\cite{Chernodub:2015gxa}, and chiral shock wave~\cite{Sen:2016jzl}.}
So far, these chiral phenomena have been mainly studied in the context of quark-gluon plasmas in heavy-ion collision experiments at Relativistic Heavy Ion Collider~(RHIC) and Large Hadron Collider~(LHC)~\cite{Kharzeev:2015znc}; see also refs.~\cite{Rybalka:2018uzh,Shovkovy:2018tks} for related works.

In this paper, we study the novel asteroseismology of the CMW of the quark number density in relativistic quark matter inside neutron stars and core-collapse supernovae and the CVW of the neutrino number density in neutrino matter at the core of supernovae.%
\footnote{One might also expect the CMW for relativistic electron matter in neutron stars and supernovae. However, the electric conductivity causes the strong damping of the CMW of the electric charge, as we discuss in appendix~\ref{app:CMW_electon_matter}.}
We call new types of oscillation modes for these chiral waves the chiral magnetic mode~(CM-mode) and chiral vortical mode~(CV-mode), respectively, in the context of asteroseismology.
We derive the dispersion relations of the CM-mode and CV-mode taking into account the chirality flipping due to the finite quark mass%
\footnote{The chirality flipping rate in electron matter in neutron stars and supernovae is computed in refs.~\cite{Grabowska:2014efa,Dvornikov:2015iua}.
In the context of the early Universe, the electron chirality flipping is also studied in refs.~\cite{Boyarsky:2020cyk,Boyarsky:2020ani}.} and diffusion.
We also estimate the frequencies and amplitudes of gravitational waves for the CM-mode and the CV-mode.
Since the frequency of the CM-mode depends on the magnetic field inside neutron stars and supernovae, we expect that the gravitational waves for the CM-mode provide a possible new probe for measuring the magnetic field strength in these compact stars. 
Furthermore, as the CM-mode can exist only in quark matter with nearly gapless quarks, the detection of the corresponding gravitational waves would suggest the existence of quark matter there.

This paper is organized as follows:
In section~\ref{sec:asteroseismology}, we briefly review the oscillation modes in the conventional asteroseismology.
In sections~\ref{sec:CM-mode_q} and \ref{sec:CV-mode}, we study the CM-mode of relativistic quark matter in neutron stars and supernovae and the CV-mode in neutrino matter at the core of supernovae, respectively. 
In section~\ref{sec:angular}, we study the angular dependence of the gravitational radiation of the CM-mode and CV-mode.
We finally conclude with discussions in section~\ref{sec:discussions}.
In this paper, we use the natural units $c=\hbar=1$.

\section{Review of the conventional asteroseismology}
\label{sec:asteroseismology}

In this section, we give a brief review of the conventional asteroseismology. We summarize the typical modes and their physical origins as well as the so-called Cowling approximation~\cite{Cowling:1941} that will be used in our analysis below.

\subsection{Classification of oscillation modes}

The oscillation modes in stars were initially classified in the context of Newtonian gravity. There are several kinds of modes such as the fundamental mode (f-mode), pressure mode (p-mode), gravity mode (g-mode), and rotational mode (r-mode)~\cite{Cowling:1941,Papaloizou:1978zz,Saio:1982}.
The f-mode and p-mode are driven by pressure, and the f-mode is a fundamental oscillation of the p-mode.
The g-mode is caused by buoyancy due to the difference in the gravitational potential, and this is why the name has the term ``gravity."
When the stars are rotating, there also exists the r-mode driven by the Coriolis force.

Asteroseismology can also be applied to compact stars such as neutron stars and supernovae in the framework of general relativity. In such a case, gravitational waves are expected to be radiated because of the seismic oscillations~\cite{McDermott:1983,Andersson:1997rn}.
In addition, there exists a mode specific to general relativity called the gravitational-wave mode (w-mode)~\cite{Kokkotas:1986,Andersson:1996,Kokkotas:1991}.
The w-mode is strongly damped with the timescale $\sim10^{-5}~$sec~\cite{Kokkotas:1999} and the frequency is independent of the structure inside stars. Since  p-, g-, and r-modes originate from the fluid oscillations of stars, they are referred to as ``fluid modes,'' while the w-mode is called ``spacetime mode.'' The typical frequencies are summarized in table~\ref{tab:modes}.

The frequencies of the modes above are, in general, complex numbers because these oscillations are damped by the emission of gravitational waves, and they are called quasi-normal modes.
\begin{table}[htb]
\begin{center}
\caption{The typical modes are shown with their physical origins and the order of magnitude of the frequencies%
~\cite{Kokkotas:1999,Andersson:2000mf}.}
\label{tab:modes}
\medskip
\begin{tabular}{|c|c|c|c|}
\hline
     & mode & physical origin & typical frequency 
    \\
    \hline\hline
    \multirow{4}{*}{fluid mode} & f-mode & pressure & $\sim10^{3}$~Hz 
    \\
    \cline{2-4}
     & p-mode & pressure & $\sim10^{3}$~Hz 
    \\
    \cline{2-4}
     & g-mode & buoyancy & $\sim10$~Hz 
    \\
    \cline{2-4}
     & r-mode & Coriolis force & $\sim1$~Hz 
    \\
    \hline
    spacetime mode & w-mode & general relativity & $\sim10^{4}$~Hz 
    \\
    \hline
\end{tabular}
\end{center}
\end{table}

\subsection{Cowling approximation}

Generally, it is hard to solve simultaneous differential equations for the fluctuations of the metric and other hydrodynamic variables to determine the frequencies of seismic oscillations.
It is then helpful to utilize the so-called Cowling approximation to simplify the differential equations.

While the Cowling approximation was first introduced in Newtonian gravity, where the recoil of the gravitational potential is ignored~\cite{Cowling:1941}, it can be extended to general relativity as well. In general relativity, one neglects the coupling between gravitational waves and the source, which makes the wave equations simpler. 
The source no longer receive the recoil by the gravitational waves and may have a different frequency~$\omega_{\rm Cowling}$ than the original frequency~$\omega_{\rm mode}$. 
The underlying assumption is that the change of the frequency is sufficiently small and gravitational waves with the same frequency as the source $(\omega_{\rm Cowling}\simeq\omega_{\rm mode})$ are emitted.
Practically, we can obtain the frequency of the gravitational waves by solving the wave equations of the source without the fluctuation of the metric in the Cowling approximation. 

It is known that the estimates of the frequencies, such as the p- and g-modes in the Cowling approximation, are within at least 20\% accuracy compared with the numerical results without the Cowling approximation%
~\cite{Yoshida:1997bf,Sotani:2020mwc}.
One can thus expect that the Cowling approximation is also valid semi-quantitatively for other modes, including the ones we will discuss in this paper.

\section{Chiral magnetic mode of quark matter in neutron stars and supernovae}
\label{sec:CM-mode_q}

We consider quark matter inside non-rotating neutron stars and supernovae in a magnetic field $\bm{B}$. 
As the typical length scale of the propagation of the CMW should be much smaller than the correlation length of the magnetic field, we can assume that the magnetic field is uniform.

In the following, we adopt the cylindrical coordinates~$\{t,r,\phi,z\}$ and take the uniform magnetic field in the $z$-axis, $\bm{B}=B\bm{e}_{\hat{z}}~(B>0)$, where $\bm{e}_{\hat{z}}$ is the normalized basis. 
The line element is then given by
\begin{align}
\label{eq:CM-metric}
    \dif s^2
    &=
    -\e^{2\lambda}\dif t^2+\e^{2\nu}\dif r^2+r^2\dif \phi^{2}+\e^{2\rho}\dif z^{2},
\end{align}
where $\lambda,\nu,\rho$ are the metric functions.
The background metric is
\begin{align}
    \bar{g}_{\alpha\beta}
    =
    \left(
    \begin{array}{cccc}
    -\e^{2\lambda} & 0 & 0 & 0 \\
    0 & \e^{2\nu} & 0 & 0 \\
    0 & 0 & r^{2} & 0 \\
    0 & 0 & 0 & \e^{2\rho}
    \end{array}
    \right).
\end{align}
The metric functions, in general, depend on $r$ and $z$ for steady and axisymmetric systems. 
However, as the typical length scale of the propagation of the CMW is also much smaller than the length scale of the variation of the metric functions, we can ignore the $z$ dependence of the metric functions.

The total metric $g_{\alpha\beta}$ is given by
\begin{align}
    g_{\alpha\beta}
    =
    \bar{g}_{\alpha\beta}+h_{\alpha\beta},
    \qquad
    |h_{\alpha\beta}|\ll1.
\end{align}
In the Cowling approximation, we ignore $h_{\alpha\beta}$ and focus only on $\bar{g}_{\alpha\beta}$ in the following discussion.

\subsection{Wave equation and dispersion relation}
Let us derive the wave equation of the CMW in quark matter of the neutron stars and supernovae.
For simplicity, we focus on the fluctuations of the number density and chiral charge density (and the fluctuation of the energy density for gravitational waves) while we ignore the coupling to the momentum density in the following discussion.

We first give a general expression of the CME~\cite{Vilenkin:1980fu,Nielsen:1983rb,Fukushima:2008xe} and the so-called chiral separation effect (CSE)~\cite{Son:2004tq,Metlitski:2005pr}.
For the $N_{\rm f}$ flavor quark field $\psi$, the quark number density and chiral charge density can be written as $n=\bar{\psi}\gamma^{0}V\psi$ and $n_5 = \bar{\psi}\gamma^{0}\gamma_{5}A\psi$, respectively, where $V=A=\bm{1}_{N_{\rm f}}$ with 
$\bm{1}_{N_{\rm f}}$ being the $N_{\rm f}\times N_{\rm f}$ identity matrix.
Introducing the quark chemical potential $\mu$ coupled to $n$ and chiral chemical potential $\mu_{5}$ coupled to $n_5$, the quark number current $\bm{j}$ and axial current $\bm{j}_{\rm 5}$ can be written as
\begin{align}
\label{eq:CME}
    &\bm{j}
    =
    N_{\rm c}{\rm tr}(VAQ)\frac{e\mu_{\rm 5}}{2\pi^{2}}\bm{B}\,,
    \\
\label{eq:CSE}
    &\bm{j}_{\rm 5}
    =
    N_{\rm c}{\rm tr}(AVQ)\frac{e\mu}{2\pi^{2}}\bm{B}\,,
\end{align}
where $N_{\rm c}$ is the number of colors and $Q$ is the quark electric charge matrix given by $Q={\rm diag}(2/3,-1/3)$ for $N_{\rm f}=2$ and $Q={\rm diag}(2/3,-1/3,-1/3)$ for $N_{\rm f}=3$.
Note that ${\bm j}$ here is not the electric current. Since ${\rm tr}(VAQ)=0$ for $N_{\rm f}=3$~\cite{Kharzeev:2010gr}, we focus on two-flavor quark matter at $\beta$ equilibrium, where, e.g., $\bar{\mu}_{\rm u}\simeq 0.80\bar{\mu}_{\rm d}$ at $T=0$. Here, $T$ is the temperature and $\bar{\mu}_{\rm q}$ for ${\rm q}={\rm u,d}$ are the up and down quark chemical potentials at equilibrium, respectively. In this case, the CMW is carried by nearly gapless up and down quarks, e.g., in the two-flavor color superconductivity (2SC)~\cite{Barrois:1977xd,Bailin:1983bm,Alford:1997zt,Rapp:1997zu} in which one of three colors does not participate in the Cooper pairing.
The coefficients of eqs.~(\ref{eq:CME}) and (\ref{eq:CSE}) are related to the chiral anomaly \cite{Adler:1969gk,Bell:1969ts,Nielsen:1983rb} and they are quantized topologically~\cite{Son:2009tf,Son:2012wh,Son:2012zy}.
Therefore, these coefficients are not affected even in curved space. 

We now consider the fluctuations of the quark number density $\delta n$ and chiral charge density $\delta n_{\rm 5}$.
The quark chemical potentials can be decomposed as
\begin{align}
    &\mu
    =
    \bar{\mu}+\delta\mu\,,
    \qquad
    \delta\mu
    =
    \frac{1}{\chi}\delta n\,,
    \\
    &\mu_{\rm 5}
    =
    \bar{\mu}_{\rm 5}+\delta\mu_{\rm 5}\,,
    \qquad
    \delta\mu_{\rm 5}
    =
    \frac{1}{\chi}\delta n_{\rm 5}\,,
\end{align}
where $\chi$ is the quark number susceptibility given by
\begin{align}
\label{eq:susceptibility_q}
\chi\equiv \frac{\del \bar{n}}{\del \bar{\mu}}=2N_{\rm c}\left(\frac{\bar{\mu}^2}{\pi^2}+\frac{T^2}{3}\right) 
\end{align}
in the ideal gas approximation.
Then the fluctuations of the currents can be expressed as
\begin{align}
\label{linearized-CME-current}
    &\delta \bm{j}
    =
    \frac{N_{\rm c} e\bm{B}}{6\pi^{2}\chi}\delta n_{\rm 5}
    -D\bm{\nabla}\delta n\,,
    \\
\label{linearized-CSE-current}
    &\delta \bm{j}_{\rm 5}
    =
    \frac{N_{\rm c} e\bm{B}}{6\pi^{2}\chi}\delta n
    -D\bm{\nabla}\delta n_{5}\,.
\end{align}
Here, we used the fact that $n$ and $n_5$ have the same susceptibility $\chi$ and diffusion coefficient $D$ in the chirally symmetric phase~\cite{Kharzeev:2010gd}.

Since we are interested in the propagation direction, we include only the $z$-derivative.
Using the relations between the bases $\bm{e}_{\alpha}$ considered here and the normalized bases $\bm{e}_{\hat{\alpha}}$,
\begin{align}
    \bm{e}_{t}
    =
    \e^{\lambda}\bm{e}_{\hat{t}},
    \qquad
    \bm{e}_{r}
    =
    \e^{\nu}\bm{e}_{\hat{r}},
    \qquad
    \bm{e}_{\phi}
    =
    r\bm{e}_{\hat{\phi}},
    \qquad
    \bm{e}_{z}
    =\e^{\rho}\bm{e}_{\hat{z}},
\end{align}
we can express the fluctuations of the currents $\delta j^{\alpha}$ and $\delta j_{\rm 5}^{\alpha}$ explicitly as
\begin{align}
\label{eq:linearized-CME-current-component}
    &\delta j^{\alpha}
    =
    \left(
    \e^{-\lambda}\delta n,
    0,
    0,
    \e^{-\rho}\frac{N_{\rm c} eB}{6\pi^2 \chi}\delta n_{\rm 5}
    -\e^{-\rho}D\del_{z}\delta n\right),
    \\
\label{eq:linearized-CSE-current-component}
    &\delta j_{\rm 5}^{\alpha}
    =
    \left(
    \e^{-\lambda}\delta n_{\rm 5},
    0,
    0,
    \e^{-\rho}\frac{N_{\rm c} eB}{6\pi^2 \chi}\delta n
    -\e^{-\rho}D\del_{z}\delta n_{5}\right).
\end{align}

We next consider the continuity equations for $n$ and $n_5$. Because of the nonzero up and down quark masses, $n_{\rm 5}$ is no longer conserved strictly while $n$ is still conserved. In the Cowling approximation, the linearized continuity equations for the quark number and chiral charge currents can be written as
\begin{align}
\label{eq:linearized-continuity-V}
    &\del_{t}\delta j^{t} + \del_{z}\delta j^{z} 
    =0\,,
    \\
\label{eq:linearized-continuity-C}
    &\del_{t}\delta j_{\rm 5}^{t} + \del_{z}\delta j_{\rm 5}^{z}
    =
    -\Gamma_{\rm flip}\delta j_{\rm 5}^{t} \,,
\end{align}
where $\Gamma_{\rm flip}$ is the chirality flipping rate for $n_5$.%
\footnote{Precisely speaking, the QCD anomaly (or the instanton effect) also contributes to the chirality flipping. While this effect is suppressed by a large power of $\Lambda_{\rm QCD}/\mu$ at sufficiently large density where the weak-coupling analysis is feasible~\cite{Schafer:2002ty}, its possible relevance is uncertain in the density region of our interest, $\mu\simeq 500~{\rm MeV}$ (see below). We here assume that, by extrapolating the above formula to the intermediate density region, the instanton effect is negligibly small compared with the quark mass effect.}
The explicit expression of $\Gamma_{\rm flip}$ will be discussed later.

We assume that the density fluctuations $\delta n,\delta n_{5}$ are written in the form of plane waves $\propto \e^{-\im(\omega t-k_{z} z)}$ with ${\rm Re}(\omega) > 0$.
Substituting eqs.~(\ref{eq:linearized-CME-current-component}) and (\ref{eq:linearized-CSE-current-component}) into eqs.~(\ref{eq:linearized-continuity-V}) and (\ref{eq:linearized-continuity-C}), we obtain
\begin{align}
    &\left(\omega+\im\e^{\lambda-\rho}D k_{z}^2\right)\delta n
    -\e^{\lambda-\rho}\frac{N_{\rm c} eB}{6\pi^2 \chi}k_{z}\delta n_{\rm 5}
    =
    0\,,
    \\
    &-\e^{\lambda-\rho}\frac{N_{\rm c} eB}{6\pi^2 \chi}k_{z}\delta n
    +\left(\omega+\im\e^{\lambda-\rho}D k_{z}^2+\im\Gamma_{\rm flip}\right)\delta n_{\rm 5}
    =
    0\,.
\end{align}
From these, we can derive the dispersion relation of the CM-mode:
\begin{align}
\label{eq:dispersion-CM}
    \omega
    =
    \sqrt{\left(V_{\rm CM}k_{z}\right)^2
    -\left(\frac{\Gamma_{\rm flip}}{2}\right)^2}
    -\im\frac{\Gamma_{\rm flip}}{2}
    -\im\e^{\lambda-\rho}Dk_{z}^2
    \equiv
    \omega_{\rm CM}\,,
\end{align}
where we have defined the speed of the CM-mode as
\begin{align}
    V_{\rm CM}
    \equiv
    \e^{\lambda-\rho}\frac{N_{\rm c} eB}{6\pi^2 \chi}\,.
\end{align}
As $\omega_{\rm CM}$ is independent of $\bar \mu_{5}$, the CMW can appear even when the system does not have the chirality imbalance at equilibrium.

As we will see later, the following condition is satisfied when the magnetic field is sufficiently strong:
\begin{align}
    V_{\rm CM}|k_{z}| \gg \frac{\Gamma_{\rm flip}}{2}.
\end{align}
In this regime, we can expand eq.~(\ref{eq:dispersion-CM}) in terms of $\Gamma_{\rm flip}$, and the resulting dispersion relation is
\begin{align}
\label{eq:dispersion-CM-expansion}
    \omega_{\rm CM}
    =
    V_{\rm CM}|k_{z}|
    -\im\frac{\Gamma_{\rm flip}}{2}
    -\im\e^{\lambda-\rho}Dk_{z}^2
    +\mathcal{O}(\Gamma_{\rm flip}^2)\,.
\end{align}
As the energy density fluctuation $\delta\varepsilon$ is proportional to $\delta n$ for $\mu \gg T$, this CM-mode causes the oscillations of the energy density and generates gravitational waves; see also  section~\ref{sec:angular}.

\subsection{Estimate of the frequency}

Let us estimate the order of magnitude of the frequency of the CM-mode~$f_{\rm CM}$. In the case of neutron stars or supernovae, the metric functions are given by
\begin{align}
\label{eq:range_wavenumber}
    \lambda
    \sim
    \rho
    \sim
    \frac{GM}{R}
    \sim
    10^{-1}\,,
\end{align}
and so $\e^{\lambda-\rho} \sim 1$, where $G$ is the gravitational constant, $M$ is the mass of the star, and $R$ is the radius.

As already mentioned, we are interested in the regime $|{\rm Re}(\omega_{\rm CM})|\gg|{\rm Im}(\omega_{\rm CM})|$.
This condition leads to the possible range of  the wavenumber as
\begin{align}
    \frac{\Gamma_{\rm flip}}{2V_{\rm CM}}
    \ll
    |k_{z}|
    \ll
    \frac{V_{\rm CM}}{D}\,,
\end{align}
where $D=\tau/3$ with $\tau$ being the relaxation time~\cite{Heiselberg:1993cr}.
Note that the condition (\ref{eq:range_wavenumber}) is more stringent than the condition of the hydrodynamic limit, $|k_{z}|\ll 2\pi/l_{\rm mfp}$, where $l_{\rm mfp}$ is the mean free path in quark matter.
Therefore, the range of the frequency of the CM-mode is
\begin{align}
\label{eq:CM-range_quark}
    \frac{\Gamma_{\rm flip}}{4\pi}
    \ll
    f_{\rm CM}
    \ll
    \frac{3V_{\rm CM}^2}{2\pi\tau}\,.
\end{align}
To obtain the explicit values of these quantities, we need to compute the chirality flipping rate~$\Gamma_{\rm flip}$ and the relaxation time~$\tau$.

\subsubsection{Chirality flipping rate}
\label{sec:chirality_flipping}

\begin{figure}[ht]
\centering
\includegraphics[width=8cm]{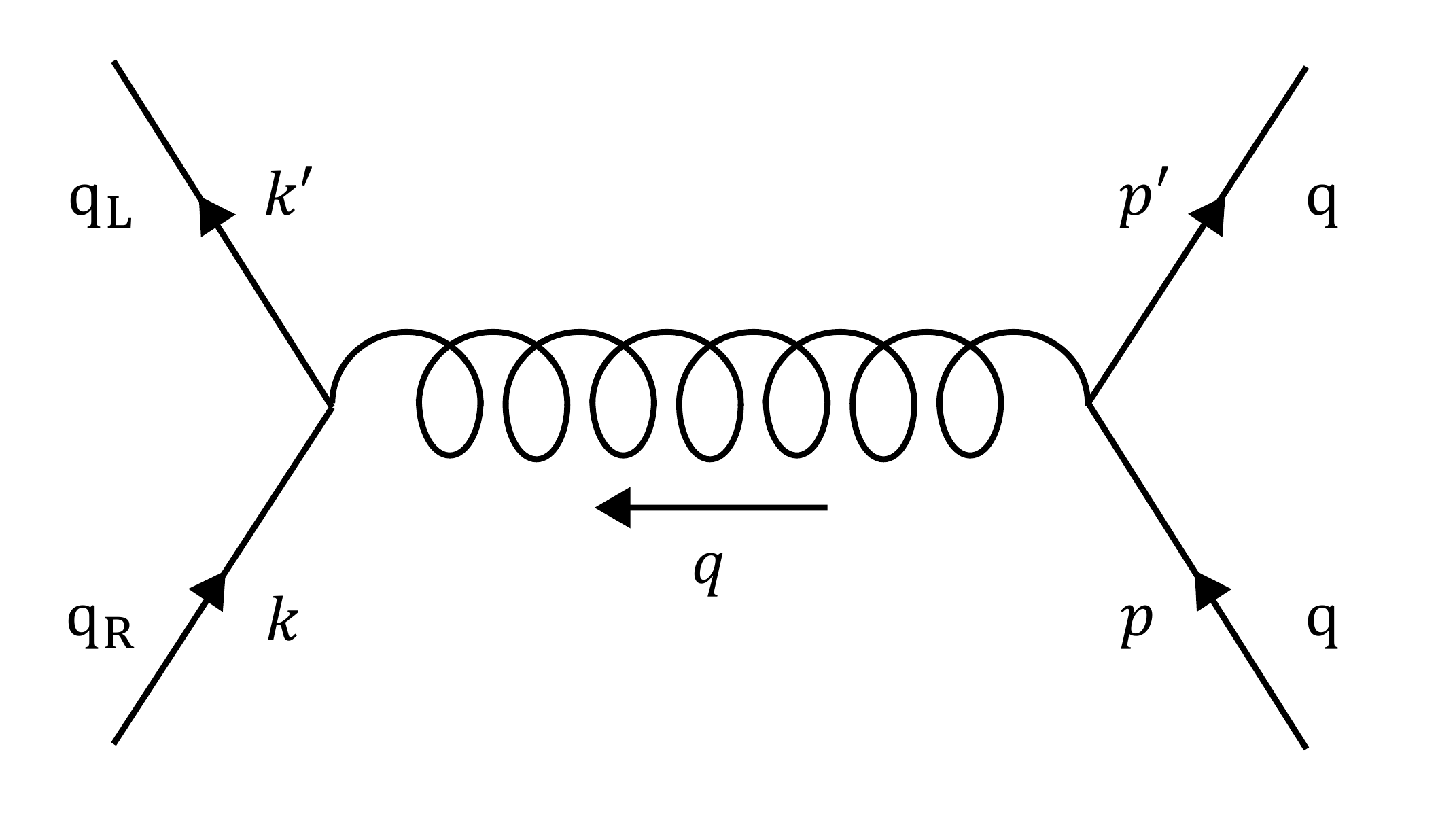}
\caption{The Feynman diagram of the chirality flipping by the quark-quark scattering.
In this figure, the right-handed quark with momentum $k$ flips to the left-handed one with momentum $k'$, while the quark with momentum $p$ does not flip the chirality.
Here, $q=k'-k$ is the energy-momentum transfer.}
\label{fig:q-q-flip-diagram}
\end{figure}

Let us compute the chirality flipping rate of quark matter in neutron stars and supernovae. Our derivation here partly follows ref.~\cite{Heiselberg:1993cr}.
Although we will eventually consider the 2SC phase as a concrete example, we here derive the expression of the chirality flipping rate for (nearly) gapless quarks with the generic number of colors, $N_{\rm c}$.
The leading-order contribution on the chirality flipping is the quark-quark scattering in figure~\ref{fig:q-q-flip-diagram}. 

The chirality flipping rate of the quark chiral charge is given as
\begin{align}
\label{eq:Gamma_flip_def_B}
    \Gamma_{\rm flip}
    \equiv
    -\frac{\dot{n}_{5}}{n_{5}} \simeq
    -\frac{\dot{n}_{5,\rm q}}{n_{5,\rm q}}
    \,,
\end{align}
where $n_{5,\rm q}$ is the quark chiral charge density for ${\rm q}={\rm u,d}$ and we have assumed that $n_{\rm 5,u} \simeq n_{\rm 5,d}$.
Below in section~\ref{sec:chirality_flipping}, we omit the index $\rm q$ for $n_{5,{\rm q}}$, $\bar \mu_{\rm q}$, etc. for notational simplicity unless otherwise stated.
To compute $\Gamma_{\rm flip}$, we use the Boltzmann equation. Introducing the distribution functions of right- and left-handed quarks as
\begin{align}
    &f_{\rm{R}}(k,t)
    =
    \frac{1}{\exp[\beta(k-\bar{\mu}-\mu_{5})]+1}\,,
    \\
    &f_{\rm{L}}(k,t)
    =
    \frac{1}{\exp[\beta(k-\bar{\mu}+\mu_{5})]+1}\,,
\end{align}
with $\beta=1/T$, we can write the chiral charge density as
\begin{align}
\label{eq:n5}
    n_{5}(t)
    &=
    N_{\rm c}\int \frac{\dif^{3}\bm{k}}{(2\pi)^{3}}\left[f_{\rm R}(k,t)-f_{\rm L}(k,t)\right]
    \nonumber \\
    &\simeq
    2N_{\rm c}\int \frac{\dif^{3}\bm{k}}{(2\pi)^{3}}\frac{\del f(k)}{\del \bar{\mu}}\mu_{5}(t)
    \nonumber \\
    &\simeq
    \frac{N_{\rm c} \bar{\mu}^{2}}{\pi^{2}}\mu_{5}(t)\,,
\end{align}
where $f$ is the equilibrium quark distribution function,
\begin{equation}
    f(k)
    =
    \frac{1}{\exp[\beta(k-\bar{\mu})]+1}\,.
\end{equation}
In eq.~(\ref{eq:n5}), we have assumed that the time dependence comes only from $\mu_{5}(t)$ and we also have used the fact that $\del f/\del \bar{\mu}$ is peaked at $k=\bar{\mu}$ for degenerate quarks.

Using the Boltzmann equations for $f_{\rm{R}}$ and $f_{\rm{L}}$,
\begin{align}
\label{eq:Boltzmann-fR}
    &\dot{f}_{\rm R}(k,t)
    =
    -C(k,t)\,,
    \\
\label{eq:Boltzmann-fL}
    &\dot{f}_{\rm L}(k,t)
    =
    +C(k,t)\,,
\end{align}
we can express $\dot{n}_{\rm5}$ as
\begin{align}
    \dot{n}_{5}(t)
    =
    -2N_{\rm c}\int \frac{\dif^{3}\bm{k}}{(2\pi)^{3}}C(k,t)\,,
\end{align}
where $C(k,t)$ is the collision integral given by
\begin{align}
    C(k,t)
    &=
    \int\frac{\dif^3 \bm{k}'}{(2\pi)^3 2k'}\frac{\dif^3 \bm{p}}{(2\pi)^3 2p}\frac{\dif^3 \bm{p}'}{(2\pi)^3 2p'}
    \frac{|M_{\rm flip}|^2}{2k}
    (2\pi)^4 \delta^{(4)}(k^{\alpha}+p^{\alpha}-k'^{\alpha}-p'^{\alpha})
    \nonumber \\
    &\times
    \left(
    f_{\rm R}(k)f(p)[1-f_{\rm L}(k')][1-f(p')]
    -f_{\rm L}(k')f(p')[1-f_{\rm R}(k)][1-f(p)]
    \right)\,,
\end{align}
where $M_{\rm flip}$ is the amplitude of the chirality flipping by the quark-quark scattering averaged over initial colors.
We note that the signs of the collision integrals in eqs.~(\ref{eq:Boltzmann-fR}) and (\ref{eq:Boltzmann-fL}) are opposite for ${f}_{\rm R}$ and ${f}_{\rm L}$ such that $\dot{f}_{\rm R} + \dot{f}_{\rm L} = 0$.

Expanding $f_{\rm{R}}$ and $f_{\rm{L}}$ to first order in $\beta \mu_{5}$,
\begin{align}
    f_{\rm{R/L}}(k,t)
    \simeq
    f(k)\pm f(k)[1-f(k)]\beta\mu_{5}(t)\,,
\end{align}
we can rewrite the chirality flipping rate (\ref{eq:Gamma_flip_def_B}) as
\begin{align}
\label{eq:Gamma_flip_simeq}
    \Gamma_{\rm flip}
    \simeq
    \frac{4\pi^{2}\beta}{\bar{\mu}^{2}}
    &\int\frac{\dif^3 \bm{k}}{(2\pi)^3 2k}\frac{\dif^3 \bm{k}'}{(2\pi)^3 2k'}\frac{\dif^3 \bm{p}}{(2\pi)^3 2p}\frac{\dif^3 \bm{p}'}{(2\pi)^3 2p'}
    |M_{\rm flip}|^2
    (2\pi)^4 \delta^{(4)}(k^{\alpha}+p^{\alpha}-k'^{\alpha}-p'^{\alpha})
    \nonumber \\
    &\times f(k)f(p)[1-f(k')][1-f(p')]\,.
\end{align}
To obtain this expression, we used
\begin{align}
     &
     f_{\rm R}(k)f(p)[1-f_{\rm L}(k')][1-f(p')]
     -
     f_{\rm L}(k')f(p')[1-f_{\rm R}(k)][1-f(p)]
     \nonumber \\
     &\simeq
     f(k)f(p)[1-f(k')][1-f(p')]\beta\mu_{5}
     +f(k')f(p')[1-f(k)][1-f(p)]\beta\mu_{5}
     \nonumber \\
     &\to
     2f(k)f(p)[1-f(k')][1-f(p')]\beta\mu_{5}\,.
 \end{align}
In the third line, we utilized the fact that the collision integral remains unchanged under the replacement of momenta $k\leftrightarrow k'$ and $p\leftrightarrow p'$ for the equilibrium distribution functions in the second term.

The scattering amplitude in eq.~(\ref{eq:Gamma_flip_simeq}) is given by
\begin{align}
    |M_{\rm flip}|^{2}
    =
    F_{\rm c}\sum_{h}|(\im g_{\rm s})^2 j^{\alpha}_{-+} D_{\alpha\beta} j ^{\beta}_{h}|^{2}\,,
\end{align}
where 
\begin{align}
F_{\rm c}\equiv \frac{1}{N_{\rm c}^2}{\rm tr}(T^a T^b){\rm tr}(T^a T^b) = \frac{N_{\rm c}^2-1}{4N_{\rm c}^2}\,,
\end{align}
with $T^a$ ($a=1,2,\cdots,N_{\rm c}^2-1$) being the generators of SU($N_{\rm c}$), $g_{\rm s}$ is the strong coupling constant, $j^{\alpha}_{-+} = \bar{u}_{-}(k')\gamma^{\alpha}u_{+}(k)$ and $j ^{\alpha}_{h} = \bar{u}_{h}(p')\gamma^{\alpha}u_{h}(p)$ are the quark currents that do and do not involve the chirality flipping, respectively (see figure~\ref{fig:q-q-flip-diagram}), with $u_{h}$ being the Dirac spinors with helicity $h$ (see appendix~\ref{app:helicity_spinor} for the explicit expressions of the spinors).
The in-medium gluon propagator $D_{\alpha\beta}$ is given by
\begin{align}
    D_{\alpha\beta}
    &=
    \frac{P_{{\rm L},\alpha\beta}}{-(q^{0})^{2}+q^{2}+\Pi_{\rm L}}
    +\frac{P_{{\rm T},\alpha\beta}}{-(q^{0})^{2}+q^{2}+\Pi_{\rm T}}\,,
\end{align}
where $q^{\alpha}=k'^{\alpha}-k^{\alpha}$ is the energy-momentum transfer, $\Pi_{\rm L}$ and $\Pi_{\rm T}$ are the longitudinal and transverse self-energies, and $q\equiv |\bm{q}|$.
The longitudinal and transverse projection operators $P_{{\rm L},\alpha\beta},P_{{\rm T},\alpha\beta}$ are defined as
\begin{align}
    &P_{{\rm T},00}=P_{{\rm T},0a}=P_{{\rm T},a0}=0\,,
    \\
    &P_{{\rm T},ab}=\delta_{ab}-\frac{q_{a}q_{b}}{q^{2}}\,,
    \\
    &P_{{\rm L},\alpha\beta}+P_{{\rm T},\alpha\beta}=\eta_{\alpha\beta}-\frac{q_{\alpha}q_{\beta}}{-(q^0)^2+q^{2}}\,,
\end{align}
with $a,b=x,y,z$ being spatial indices.

We decompose the square of the amplitude of the chirality flipping by the quark-quark scattering,  $|M_{\rm flip}|^{2}$, into the following form:
\begin{align}
    |M_{\rm flip}|^{2}
    =
    |M_{\rm flip,L}|^2
    +|M_{\rm flip,T}|^2
    +2|M_{\rm flip,cross}|^2,
\end{align}
where 
\begin{align}
    &|M_{\rm flip,L}|^2
    =
    F_{\rm c}\sum_{h}\left|\frac{j_{-+}^{0}j_{h}^{0}}{q^{2}+\Pi_{\rm L}}\right|^2,
    \qquad
    |M_{\rm flip,T}|^2
    =
    F_{\rm c}\sum_{h}\left|\frac{\bm{j}_{-+,\rm T}\cdot \bm{j}_{h,\rm T}}{-(q^{0})^2+q^{2}+\Pi_{\rm T}}\right|^2, \nonumber \\
    &|M_{\rm flip,cross}|^2 = F_{\rm c}\sum_{h}\left|\frac{j_{-+}^{0}j_{h}^{0}}{q^{2}+\Pi_{\rm L}} \cdot \frac{\bm{j}_{-+,\rm T}\cdot \bm{j}_{h,\rm T}}{-(q^{0})^2+q^{2}+\Pi_{\rm T}}\right|\,,
\end{align}
with $\bm{j}_{\rm T}$ being the transverse component of the current with respect to $\bm{q}$. In the following, we simply focus on the contribution of $|M_{\rm flip,L}|^2$ and $|M_{\rm flip,T}|^2$, since the integral of $|M_{\rm flip,cross}|^2 \propto \cos{\varphi}$ over $\varphi$ vanishes, where $\varphi$ is the angle between $\bm{j}_{-+,\rm T}$ and $\bm{j}_{h,\rm T}$. 

We set $\bm{k}$ to be along the $z$-axis.
For simplicity of calculation, we focus on the special case $\theta_{pk} = 0$, where $\theta_{pk}$ is the angle between $\bm{p}$ and $\bm{k}$, since the resulting parameter dependence of the chirality flipping rate does not depend on the particular choice of $\theta_{pk}$.
As the integration over $\varphi$ gives a constant and does not change the physical parameter dependence of the chirality flipping, we also assume that the scattering occurs in the same plane, namely,  $\varphi=0$.
In this case, we can express the currents with and without the chirality flipping as
\begin{align}
\label{eq:general_currents_flip}
    &j_{-+}^{\alpha}
    \simeq
    -\frac{g_{\rm s}m}{\sqrt{kk'}}
    \left(
    (k+k')\sin{\frac{\theta_{kk'}}{2}}, \
    q^{0}\cos{\frac{\theta_{kk'}}{2}}, \
    \im q^{0}\cos{\frac{\theta_{kk'}}{2}},\
    -q^{0}\sin{\frac{\theta_{kk'}}{2}}
    \right),
    \\
\label{eq:general_currents_no_flip}
    &j_{h}^{\alpha}
    \simeq
    2g_{\rm s}\sqrt{pp'}
    \left(
    \cos{\frac{\theta_{pp'}}{2}}, \ 
    \sin{\frac{\theta_{pp'}}{2}}, \ 
    \im h\sin{\frac{\theta_{pp'}}{2}}, \ 
    \cos{\frac{\theta_{pp'}}{2}}
    \right).
\end{align}
Here, we have expanded $j_{-+}^{\alpha}$ and $j_{h}^{\alpha}$ up to the leading first order and zeroth order of the quark mass $m$, respectively. 

We first consider the longitudinal contribution.
Using eqs.~(\ref{eq:general_currents_flip}) and (\ref{eq:general_currents_no_flip}), the longitudinal amplitude can be written as
\begin{align}
    |M_{\rm flip,L}|^{2}
    &=
    2F_{\rm c}\left|\frac{j_{-+}^{0}j_{+}^0}{q^2+\Pi_{\rm L}}\right|^2
    \nonumber \\
    &=
    32\pi^{2}F_{\rm c}\alpha_{\rm s}^2 m^{2}\frac{pp'}{kk'}\frac{(k+k')^2}{|q^2+\Pi_{\rm L}|^2}
    (1-\cos{\theta_{kk'}})(1+\cos{\theta_{pp'}})\,.
\end{align}
Here, the factor 2 comes from the summation over $h$ with $j_{+}^0=j_{-}^{0}$, and $\alpha_{\rm s}\equiv g_{s}^{2}/(4\pi)$. 
We note that the factor $(1-\cos{\theta_{kk'}})$ is characteristic of the chirality flipping~\cite{Schafer:2001za} in the so-called high density effective theory near the Fermi surface~\cite{Hong:1998tn,Hong:1999ru}.
We rewrite the delta function as
\begin{align}
    \delta^{(4)}(k+p-k'-p')
    =
    \int\dif q^{0}\dif^{3}\bm{q}
    \delta(k'-k-q^{0})\delta(p'-p+q^{0})
    \delta^{(3)}(\bm{k}'-\bm{k}-\bm{q})\delta^{(3)}(\bm{p}'-\bm{p}+\bm{q}),
\end{align}
and introduce the functions $S_{1,\rm L}$ and $S_{2,\rm L}$:
\begin{align}
    S_{1,\rm L}(q^{0},\bm{q})
    &\equiv
    \int \frac{\dif^3 \bm{k}}{(2\pi)^3}\frac{\dif^3 \bm{k}'}{(2\pi)^3}
    \frac{(k+k')^{2}}{k^2 k'^2}(1-\cos{\theta_{kk'}})
    \nonumber \\
    &\quad\times(2\pi)^4 \delta(k'-k-q^{0})\delta^{(3)}(\bm{k}'-\bm{k}-\bm{q})
    f(k)[1-f(k')]\,,
    \\
    S_{2,\rm L}(q^{0},\bm{q})
    &\equiv
    \int \frac{\dif^3 \bm{p}}{(2\pi)^3}\frac{\dif^3 \bm{p}'}{(2\pi)^3}
    (1+\cos{\theta_{pp'}})
    \nonumber \\
    &\quad\times(2\pi)^4 \delta(p'-p+q^{0})\delta^{(3)}(\bm{p}'-\bm{p}+\bm{q})
    f(p)[1-f(p')]\,.
\end{align}
Using these functions, we can express the longitudinal contribution to the chirality flipping rate as
\begin{align}
\label{eq:CFR_qq_l}
    \Gamma_{\rm flip,L}
    &\simeq
    F_{\rm c}\frac{\alpha_{\rm s}^2 m^2 \beta}{2\bar{\mu}^2}
    \int \dif q^{0}\dif^3 \bm{q} 
    \frac{1}{|q^2+\Pi_{\rm L}|^2}
    S_{1,\rm L}S_{2,\rm L}\,.
\end{align}

We assume that $q^{0}(\sim T)\ll q(\sim q_{\rm IR})\ll k(\sim \bar{\mu})$, where $q_{\rm IR}$ is the IR cutoff of the momentum.%
\footnote{Since the momentum $q$ around the IR cutoff has the dominant contribution to the integral over $\bm{q}$, we can assume that $q\sim q_{\rm IR}$.}
In this case, the gluon self-energies are expressed as
\begin{align}
    \Pi_{\rm L}\simeq q_{\rm s}^2\,,
    \qquad
    \Pi_{\rm T}\simeq \im\frac{\pi}{4}q_{\rm s}^2\frac{q^{0}}{q}\,,
\end{align}
where $q_{\rm s}$ is the Debye screening mass of the gluon defined by $q_{\rm s}^2\equiv N_{\rm f}g_{\rm s}^2\bar{\mu}^2/\pi^2$.
Then we can integrate over $\bm{k}$ and $\bm{k}'$ in $S_{1,\rm L}$ as
\begin{align}
    S_{1,\rm L}(q^{0},\bm{q})
    &=
    \int \frac{\dif^3 \bm{k}}{(2\pi)^3}
    \frac{(k+k')^{2}}{k^2 k'^2}(1-\cos{\theta_{kk'}})
    \nonumber \\
    &\quad\times\left.(2\pi)\delta(k'-k-q^{0})
    f(k)[1-f(k')]\right|_{\bm{k}'=\bm{k}+\bm{q}}
    \nonumber \\
    &\simeq
    \frac{1}{2\pi q}
    \int_{k>q^{0}}\dif k \dif\cos{\theta_{kq}}\frac{(k+k')^2}{k'^2}
    (1-\cos{\theta_{kk'}})
    \nonumber \\
    &\quad\times\left.\delta\left(\cos{\theta_{kq}-\frac{q^0}{q}+\frac{q}{2k}}\right)
    f(k)[1-f(k')]
    \right|_{\bm{k}'=\bm{k}+\bm{q}}
    \nonumber \\
    &\simeq
    \frac{1}{4\pi q}\int_{k>q^{0}}\dif k\left(\frac{2k+q^{0}}{k+q^{0}}\right)^{2}\left(\frac{q}{k}\right)^{2}
    \delta(k-\bar{\mu})\left(T+\frac{q^{0}}{2}\right)
    \nonumber \\
    &\simeq
    \frac{q}{\pi \bar{\mu}^2}\left(T+\frac{q^{0}}{2}\right).
\end{align}
Here, we have used the following relations:
\begin{align}
    &\delta(|\bm{k}+\bm{q}|-k-q^{0})
    \simeq
    \frac{1}{q}\delta\left(\cos{\theta_{kq}}-\frac{q^0}{q}+\frac{q}{2k}\right)
    \theta(k+q^{0})\,,
    \\
    &\cos{\theta_{kk'}}
    =
    \frac{k+q\cos{\theta_{kq}}}{\sqrt{k^2 +q^2 +2kq\cos{\theta_{kq}}}}\,,
    \\
    &f(k)[1-f(k+q^{0})]
    \simeq
    \delta(k-\bar{\mu})\left(T+\frac{q^{0}}{2}\right)\,.
\end{align}
In the same way, we can calculate $S_{2,\rm L}$ as
\begin{align}
    S_{2,\rm L}(q^{0},\bm{q})
    &\simeq
    \frac{\bar{\mu}^2}{\pi q}\left(T-\frac{q^{0}}{2}\right).
\end{align}

Substituting $S_{1,\rm L}$ and $S_{2,\rm L}$ into eq.~(\ref{eq:CFR_qq_l}), the longitudinal contribution to the chirality flipping rate can be obtained as
\begin{align}
\label{eq:CFR_l_parameter}
    \Gamma_{\rm flip,L}
    &\simeq
    F_{\rm c}\frac{2\alpha_{\rm s}^2 m^2 \beta}{\pi\bar{\mu}^2}
    \int \dif q^{0}\dif q
    \frac{q^2}{|q^2+\Pi_{\rm L}|^2}
    \left[T^2-\left(\frac{q^{0}}{2}\right)^2\right]
    \nonumber \\
    &\simeq
    F_{\rm c}\frac{2\alpha_{\rm s}^2 m^2 \beta}{\pi\bar{\mu}^2 q_{\rm s}}
    \int_{0}^{T}\dif q^{0}\left[T^2-\left(\frac{q^{0}}{2}\right)^2\right]
    \int_{0}^{1}\frac{\dif \zeta}{(1+\zeta^2)^2}
    \nonumber \\
    &\simeq
    F_{\rm c}\frac{11(\pi+2)}{48\pi}\frac{\alpha_{\rm s}^2 m^2}{\bar{\mu}^2 q_{\rm s}}T^2\,.
\end{align}
In the second line, we defined $\zeta\equiv q_{\rm s}/q$.
The temperature dependence of the chirality flipping rate is consistent with the behavior of the Fermi liquid~\cite{Lihshitz:1980} and is different from that of the Rutherford scattering in refs.~\cite{Grabowska:2014efa,Dvornikov:2015iua}.
This difference comes from their assumption that the proton is so heavy that its recoil is negligible, which leads to the isoenergetic (or $q^0=0$) scattering.
In that case, since the contribution of the energy width $T$ of the proton is neglected, the temperature dependence of $\Gamma_{\rm flip,L}$ changes from $\sim T^2$ to $\sim T$.

We next evaluate the transverse contribution to the chirality flipping rate. The amplitude is given by
\begin{align}
    |M_{\rm flip,T}|^{2}
    =
    F_{\rm c}\left|\frac{\bm{j}_{-+,{\rm T}}\cdot\bm{j}_{+,{\rm T}}}{-(q^0)^2+q^2+\Pi_{\rm T}}\right|^2
    +F_{\rm c}\left|\frac{\bm{j}_{-+,{\rm T}}\cdot\bm{j}_{-,{\rm T}}}{-(q^0)^2+q^2+\Pi_{\rm T}}\right|^2.
\end{align}
Though the expressions of the two terms are different, their parameter dependence is the same, and we focus only on the first term.
Using  eqs.~(\ref{eq:general_currents_flip}) and (\ref{eq:general_currents_no_flip}), the transverse amplitude can be written as
\begin{align}
\label{eq:amplitude_qq_t}
    |M_{\rm flip,T}|^{2}
    &\simeq
    F_{\rm c}\left|\frac{j^{a}_{-+}\left[\delta_{ab}-(q_{a}q_{b}/q^2)\right]j_{+}^{b}}{-(q^0)^2+q^2+\Pi_{\rm T}}\right|^2
    \nonumber \\
    &=
    F_{\rm c}\left|\frac{j^{a}_{-+}j_{+}^{a}-\left(q^{0}/q\right)^2 j^{0}_{-+} j_{+}^{0}}{-(q^0)^2+q^2+\Pi_{\rm T}}\right|^2
    \nonumber \\
    &=
    16\pi^2 F_{\rm c}\alpha_{\rm s}^2 m^2\frac{pp'}{kk'}
    \frac{\left[q^{0}-(k+k')\left(q^{0}/q\right)^2\right]^2}{|-(q^0)^2+q^2+\Pi_{\rm T}|^2}
    (1-\cos{\theta_{kk'}})(1+\cos{\theta_{pp'}})\,.
\end{align}
In the second line, we have used the current conservation, $\bm{q}\cdot\bm{j}=q^{0}j^{0}$.

We decompose the transverse contribution to the chirality flipping rate into three parts:
\begin{align}
    \Gamma_{\rm flip,T}
    =
    \Gamma_{\rm flip,T}^{(1)}
    +
    \Gamma_{\rm flip,T}^{(2)}
    +
    \Gamma_{\rm flip,T}^{(3)}\,,
\end{align}
where
\begin{align}
    \Gamma_{\rm flip,T}^{(1)}
    &\simeq
    F_{\rm c}\frac{\alpha_{\rm s}^2 m^2 \beta}{4\bar{\mu}^2}
    \int \dif q^{0}\dif^3 \bm{q} 
    \frac{(q^{0})^2}{|-(q^0)^2+q^2+\Pi_{\rm T}|^2}
    S_{1,\rm T}^{(1)}S_{2,\rm T}\,,
    \nonumber \\
    \Gamma_{\rm flip,T}^{(2)}
    &\simeq
    F_{\rm c}\frac{\alpha_{\rm s}^2 m^2 \beta}{4\bar{\mu}^2}
    \int \dif q^{0}\dif^3 \bm{q} 
    \frac{-2q^{0}\left(q^{0}/q\right)^2}{|-(q^0)^2+q^2+\Pi_{\rm T}|^2}
    S_{1,\rm T}^{(2)}S_{2,\rm T}\,,
    \nonumber \\
    \Gamma_{\rm flip,T}^{(3)}&\simeq
    F_{\rm c}\frac{\alpha_{\rm s}^2 m^2 \beta}{4\bar{\mu}^2}
    \int \dif q^{0}\dif^3 \bm{q} 
    \frac{\left(q^{0}/q\right)^4}{|-(q^0)^2+q^2+\Pi_{\rm T}|^2}
    S_{1,\rm T}^{(3)}S_{2,\rm T}\,.
\end{align}
The functions $S_{1,\rm T}^{(i)}$ ($i=1,2,3$) and $S_{2,\rm T}$ are defined and computed as
\begin{align}
    S_{1,\rm T}^{(i)}(q^{0},\bm{q})
    &\equiv
    \int \frac{\dif^3 \bm{k}}{(2\pi)^3}\frac{\dif^3 \bm{k}'}{(2\pi)^3}
    \frac{(k+k')^{i-1}}{k^2 k'^2}(1-\cos{\theta_{kk'}})
    \nonumber \\
    &\quad\times(2\pi)^4 \delta(k'-k-q^{0})\delta^{(3)}(\bm{k}'-\bm{k}-\bm{q})
    f(k)[1-f(k')]
    \nonumber \\
    &\simeq
    \frac{q}{4\pi\bar{\mu}^{5-i}}\left(T+\frac{q^{0}}{2}\right)\,,
    \\
    S_{2,\rm T}(q^{0},\bm{q})
    &\equiv
    S_{2,\rm L}(q^{0},\bm{q})\,.
\end{align}

In the same way as the longitudinal case, we can obtain the parameter dependence as
\begin{align}
    \Gamma_{\rm flip,T}^{(1)}
    \sim
    F_{\rm c}\frac{\alpha_{\rm s}^2 m^2}{\bar{\mu}^4 q_{\rm s}^{2/3}}T^{\frac{11}{3}}\,, \quad 
    \Gamma_{\rm flip,T}^{(2)}
    \sim
    -F_{\rm c}\frac{\alpha_{\rm s}^2 m^2}{\bar{\mu}^3 q_{\rm s}^{2}}T^{4}\,, \quad
    \Gamma_{\rm flip,T}^{(3)}
    \sim
    F_{\rm c}\frac{\alpha_{\rm s}^2 m^2}{\bar{\mu}^2 q_{\rm s}^{10/3}}T^{\frac{13}{3}}\,.
\end{align}
From the ratios between the longitudinal and transverse contributions, such as
\begin{align}
    \frac{\Gamma_{\rm flip,T}^{(1)}}{\Gamma_{\rm flip,L}}
    \sim
    10^{-11}\left(\frac{T}{10^{6}~{\rm K}}\right)^{\frac{5}{3}},
    \qquad
    \frac{\Gamma_{\rm flip,T}^{(3)}}{\Gamma_{\rm flip,L}}
    \sim
    10^{-14}\left(\frac{T}{10^{6}~{\rm K}}\right)^{\frac{7}{3}},
\end{align}
we find that the transverse contribution is suppressed when $T\lesssim 10^{11}~{\rm K}$ and the longitudinal contribution is dominant in our setup ($T\sim 10^{6}\text{--}10^{11}~{\rm K}$).
This fact can be understood from the high density effective theory based on the systematic expansion of $T/\mu$~\cite{Hong:1998tn,Hong:1999ru}, where the chirality flipping is caused by the temporal component of the gauge field $A^{0}$~\cite{Schafer:2001za}, and only the longitudinal propagator is relevant. 

We now consider the case of the 2SC phase and focus on the nearly gapless unpaired quark whose energy measured from the Fermi energy is much smaller than the color superconducting gap $\Delta(\sim 10^{2}~{\rm MeV})$~\cite{Alford:1997zt}.
Since the unpaired quarks have only one color out of three, the quark-quark scattering does not contribute to the chirality flipping at tree level.
The other possible candidates for the chirality flipping due to the quark-quark scattering are the QCD process at second order in $\alpha_{\rm s}$ and the QED process at first order in $\alpha_{\rm e}(\equiv e^2/(4\pi))$.
The chirality flipping rate in the former QCD process may be parametrically estimated by multiplying eq.~(\ref{eq:CFR_l_parameter}) by $\alpha_{\rm s}^2$ as
\begin{align}
    \Gamma_{\rm flip,QCD}^{(\rm 2SC)}
     &\sim
     \frac{\alpha_{\rm s}^4 m^2}{\bar{\mu}^2 q_{\rm s}}T^2\,,
\end{align}
while that in the latter QED process is obtained with the replacement, $F_{\rm c}\to 1$ and $g_{\rm s}\to e~(\alpha_{\rm s}\to\alpha_{\rm e})$ in the above discussion:
\begin{align}
\label{eq:CFR_2SC}
    \Gamma_{\rm flip,QED}^{(\rm 2SC)}\sim
    \frac{\alpha_{\rm e}^2 m^2}{\bar{\mu}^2 q_{\rm e}}T^2\,,
\end{align}
where $q_{\rm e}$ is the Debye screening mass of the photon given by $q_{\rm e}^2 \equiv 5e^2\bar{\mu}^2/(9\pi^2)$.
Comparing these two processes, e.g., at $\bar{\mu}\simeq 
500~{\rm MeV}$ and $\alpha_{\rm s}\simeq 0.7$, we find that the QCD process is dominant in our setup:
\begin{align}
    \frac{\Gamma_{\rm flip,QCD}^{(\rm 2SC)}}{\Gamma_{\rm flip,QED}^{(\rm 2SC)}}
    \sim
    \frac{\alpha_{\rm s}^4 q_{\rm e}}{\alpha_{\rm e}^2 q_{\rm s}}
    \sim
    10^{2}\,.
\end{align}

\subsubsection{Relaxation time of quark matter}

The relaxation time in quark matter was estimated in ref.~\cite{Heiselberg:1993cr} and has two parts originating from the longitudinal and transverse gluon self-energies:
\begin{align}
\label{eq:tau_1st_order}
    \begin{array}{ll}
    \displaystyle
    \frac{1}{\tau_{\rm L}}\sim F_{\rm c}\alpha_{\rm s}^2 \frac{T^2}{q_{\rm s}} & (\text{longitudinal contribution})\,,
    \vspace{1em}
    \\
    \displaystyle
    \frac{1}{\tau_{\rm T}}\sim F_{\rm c}\alpha_{\rm s}^2 \frac{T^{5/3}}{q_{\rm s}^{2/3}} & (\text{transverse contribution})\,.
    \end{array}
\end{align}
While the longitudinal contribution is consistent with the typical Fermi-liquid behavior $\sim T^2$~\cite{Lihshitz:1980}, the transverse one has the different temperature dependence as the Landau damping results in the IR cutoff like $q_{\rm IR}\sim (q_{\rm s}^2 T)^{1/3}$ instead of $q_{\rm IR}\sim q_{\rm s}$.
In our setup ($T\sim 10^{6}\text{--}10^{11}~\rm K$), the transverse contribution is dominant.
Similarly to the chirality flipping rate above, the relaxation time in the 2SC phase is dominated by the QCD process and is given by multiplying the transverse contribution in eq. (\ref{eq:tau_1st_order}) by $\alpha_{\rm s}^2$ as
\begin{align}
\label{eq:tau_2SC}
    \frac{1}{\tau^{(\rm 2SC)}_{\rm QCD}}\sim\alpha_{\rm s}^4\frac{T^{5/3}}{q_{\rm s}^{2/3}}\,.
\end{align}

\subsubsection{Result}

Now we estimate the order of magnitude of the frequency of the CM-mode in the 2SC phase by substituting eqs.~(\ref{eq:CFR_2SC}) and (\ref{eq:tau_2SC}) into eq.~(\ref{eq:CM-range_quark}).
Here, the quark chemical potential coupled to the quark number density $n$ will be denoted by $\bar{\mu}$ again. 
Using $m_{\rm q}\simeq 3\text{--}5~{\rm MeV},~\bar{\mu}\simeq 
500~{\rm MeV},~\alpha_{\rm s}\simeq 0.7$, we obtain
\begin{align}
\label{eq:CM_freq_quark}
    10^{4}\,{\rm Hz}
    \left(\frac{T}{10^{6}\,{\rm K}}\right)^{2}
    \ll
    f_{\rm CM}
    \ll
    10^{6}\,{\rm Hz}
    \left(\frac{B}{10^{18}\,{\rm Gauss}}\right)^{2}
    \left(\frac{T}{10^{6}\,{\rm K}}\right)^{5/3}.
\end{align}
Since the power of the temperature of the upper limit is smaller than that of the lower limit, the CM-mode would not exist in high temperature ($T\gtrsim 10^{7}~\rm K$) environments such as supernovae.

From the dispersion relation~(\ref{eq:dispersion-CM-expansion}), the order of magnitude of the frequency of the CM-mode is expressed as
\begin{align}
\label{eq:frequency-CM-parameter}
    f_{\rm CM}
    \sim
    10^{5}\,{\rm Hz}
    \left(\frac{B}{10^{18}~\rm Gauss}\right)
    \left(\frac{\bar{\mu}}{500\,{\rm MeV}}\right)^{-2}
    \left(\frac{k}{10^{-4}\,{\rm /cm}}\right)\,,
\end{align}
where we ignored the second term in eq.~(\ref{eq:susceptibility_q}) for $T/\mu\ll1$.
Since the CMW cannot exist in electron matter (see appendix \ref{app:CMW_electon_matter}) and the frequency of the CM-mode can be distinguished from the conventional modes, the CM-mode provides a possible new probe for quark matter in neutron stars.
Also, as eq.~(\ref{eq:frequency-CM-parameter}) depends on the magnetic field, possible detection of this mode would provide information on the magnetic field inside neutron stars.

\subsection{Estimate of the amplitude}
We estimate the amplitude of the gravitational wave of the CM-mode.
Generally, the characteristic amplitude is given in terms of the distance to a source $d$, the released energy as gravitational waves $E_{\rm GW}$, and the frequency $f$ regardless of the details of the production mechanism~\cite{Andersson:1996pn}:
\begin{align}
\label{eq:amp-formula}
    h
    \sim
    \frac{1}{d}\sqrt{\frac{GE_{\mathrm{GW}}}{f}}\,.
\end{align}
Using this formula, the amplitude of the gravitational wave with the typical frequency of the CM-mode is given by
\begin{align}
\label{eq:CM_amp_quark}
        h_{\rm CM}
    \sim
    10^{-22}
    \left(\frac{E_{\rm GW}}{10^{44}\,{\rm erg}}\right)^{1/2}
    \left(\frac{f_{\rm CM}}{10^{5}\,{\rm Hz}}\right)^{-1/2}
    \left(\frac{d}{10\,{\rm kpc}}\right)^{-1}.
\end{align}
Here, considering possible sources of the CM-mode in neutron stars and supernovae, $E_{\rm GW}$ and $d$ are written with respect to $\sim 1$\% of the energy released by a giant flare in a neutron star ($\sim 10^{46}\,{\rm erg}$) and about the radius of our galaxy $\sim 10\,{\rm kpc}$, respectively.

\section{Chiral vortical mode of neutrino matter in supernovae}
\label{sec:CV-mode}

When a gravitational collapse begins and the density of the core increases, the electron capture ${\rm p}+\e^{-}_{\rm L}\to {\rm n}+\nu_{\rm L}$ occurs to lower the Fermi energy of the electrons. This process described by the weak interaction violates the parity symmetry and generates only left-handed neutrinos. Since the typical timescale of the neutrino diffusion is larger than the free-fall timescale, the neutrinos are trapped in the core of supernovae~\cite{Sato:1974,Kotake:2005zn}. Due to the left-handedness of neutrinos, the neutrino matter there is the ``chiral matter'' with chirality imbalance~\cite{Yamamoto:2015gzz}.

We consider the neutrino matter of supernovae rotating with the angular velocity $\bm{\Omega}$.
We use the cylindrical coordinates $\{t,r,\phi,z\}$ and orient the rotational axis along the $z$-axis, $\bm{\Omega}=\Omega\bm{e}_{\hat{z}}~ (\Omega>0)$.
The relation of the angular coordinate between the inertial and the corotating frame is $\psi=\phi-\Omega t$, where $\psi$ is the angular coordinate of the corotating frame.
In the corotating frame, the small line element is given by
\begin{align}
\label{eq:cvw-metric}
    \dif s^2
    &=
    -\e^{2\lambda}\dif t^2+\e^{2\nu}\dif r^2+r^2\dif \psi^{2}+\e^{2\rho}\dif z^{2}
    \nonumber \\
    &\simeq
    -\e^{2\lambda}\dif t^2-2r^{2}\Omega \dif t\dif \phi+\e^{2\nu}\dif r^2+r^2\dif \phi^{2}+\e^{2\rho}\dif z^{2}.
\end{align}
In the second line, we have ignored $\mathcal{O}(\Omega^{2})$.
For the same reason mentioned in section~\ref{sec:CM-mode_q}, the metric functions are independent of $z$. We also ignore the fluctuation of the metric in the Cowling approximation and focus on the background metric, 
\begin{align}
    \bar{g}_{\alpha\beta}
    =
    \left(
    \begin{array}{cccc}
        -\e^{2\lambda} & 0 & -r^{2}\Omega & 0 \\
        0 & \e^{2\nu} & 0 & 0 \\
        -r^{2}\Omega & 0 & r^{2} & 0 \\
        0 & 0 & 0 & \e^{2\rho}
    \end{array}
    \right)\,.
\end{align}

\subsection{Wave equation and dispersion relation}

We derive the wave equation of the CVW%
~\cite{Jiang:2015cva} in the core of supernovae.
The CVE of the left-handed neutrinos is%
~\cite{Vilenkin:1979ui,Son:2009tf,Landsteiner:2011cp}
\begin{align}
\label{eq:CVE}
    \bm{j}_{\nu}
    =
    -\left(\frac{\mu_{\nu}^{2}}{4\pi^{2}}+\frac{T^{2}}{12}\right)\bm{\Omega}\,,
\end{align}
where $\mu_{\nu}$ is the neutrino chemical potential and $T$ is the temperature. The coefficient of the first term in eq.~(\ref{eq:CVE}) is also fixed by the chiral anomaly~\cite{Son:2009tf} and it does not receive corrections in curved space.%
\footnote{Equation~(\ref{eq:CVE}) has also corrections proportional to the fermion mass and scalar curvature for massive fermions in curved space~\cite{Flachi:2017vlp}. However, as we are interested in nearly massless neutrinos, these corrections are negligibly small.}

In the presence of the fluctuation of the neutrino density $\delta n_{\nu}$, the neutrino chemical potential is expressed as
\begin{align}
    \mu_{\nu}
    =
    \bar{\mu}_{\nu}+\delta \mu_{\nu}\,,
    \qquad
    \delta\mu_{\nu}
    =
    \frac{1}{\chi_{\nu}}\delta n_{\nu}\,,
\end{align}
where $\chi_{\nu}$ is the number susceptibility of neutrino matter. 
Accordingly, the fluctuation of the current, $\delta \bm{j}_{\nu}$, is given by
\begin{align}
    \delta \bm{j}_{\nu}
    =
    -\frac{\bar{\mu}_{\nu}}{2\pi^2 \chi_{\nu}}\bm{\Omega}\delta n_{\nu}
    -D_{\nu}\bm{\nabla}\delta n_{\nu}\,,
\end{align}
where $D_{\nu}$ is the neutrino diffusion coefficient.
We focus on the propagation direction and include only the $z$-derivative.
Then $\delta j_{\nu}^{\alpha}$ is written as
\begin{align}
    \delta j_{\nu}^{\alpha}
    =
    \left(\e^{-\lambda}\delta n_{\nu},
    0,
    0,
    -\e^{-\rho}\frac{\bar{\mu}_{\nu}\Omega}{2\pi^2\chi_{\nu}}\delta n_{\nu}
    -\e^{-\rho}D_{\nu}\del_{z}\delta n_{\nu}
    \right)\,.
\end{align}

In the Cowling approximation, the linearized continuity equation is given by
\begin{align}
  \label{eq:continuity-nu}
  \del_t \delta j_{\nu}^{t} + \del_z \delta j_{\nu}^{z}
  =
  0\,.
\end{align}
Assuming that $\delta n_{\nu}\propto \e^{-\im(\omega t-k_{z}z)}$, eq.~(\ref{eq:continuity-nu}) becomes
\begin{align}
    \left(\omega
    +\e^{\lambda-\rho}\frac{\bar{\mu}_{\nu}\Omega}{2\pi^2\chi_{\nu}}k_{z}
    +\im \e^{\lambda-\rho}D_{\nu}k_{z}^2\right)\delta n_{\nu}
    =
    0\,.
\end{align}
Therefore, we arrive at the dispersion relation of the CV-mode,
\begin{align}
\label{eq:dispersion-CV}
    \omega = -V_{\rm CV}k_{z}-\im\e^{\lambda-\rho}D_{\nu}k_{z}^2
    \equiv \omega_{\rm CV},
\end{align}
where 
\begin{equation}
    V_{\rm CV} \equiv \e^{\lambda-\rho}\frac{\bar{\mu}_{\nu}\Omega}{2\pi^2\chi_{\nu}}
\end{equation}
is the speed of the CV-mode. Note that the propagation of the CV-mode requires the presence of nonzero left-handed neutrino chemical potential (or chirality imbalance) at equilibrium unlike the CM-mode.

\subsection{Possible existence of the CV-mode in the core of supernovae}

We estimate the frequency of the CV-mode at the core of supernovae.
In the same way as section~\ref{sec:CM-mode_q}, we set $\e^{\lambda-\rho}\sim 1$.
The wavelength has to be shorter than the radius $R$, and the condition $|{\rm Re}(\omega_{\rm CV})|\gg |{\rm Im}(\omega_{\rm CV})|$ is necessary for the propagation of the CV-mode:
\begin{align}
    \frac{2\pi}{R}\ll |k_{z}| \ll \frac{V_{\rm CV}}{D_{\nu}}\,,
\end{align}
where $D_{\nu} \sim l_{\rm mfp}/3$ with the assumption that the relaxation timescale is comparable to the mean free path, $\tau \sim l_{\rm mfp}$.
Assuming that neutrino matter is an ideal neutrino gas, the susceptibility is given as
\begin{align}
    \chi_{\nu}
    \equiv
    \frac{\del n_{\nu}}{\del\bar{\mu}_{\nu}}=
    \frac{\bar{\mu}_{\nu}^2}{2\pi^2}
    +\frac{T^2}{6}\,.
\end{align}

We now estimate the upper and lower limits of the wavenumber.
Taking the radius of the core of supernovae, $R\sim10\,\mathrm{km}$, neutrino chemical potential $\bar{\mu}_{\nu}\sim10^2\,{\rm MeV}$, and  temperature $T\sim10\,{\rm MeV}$, the lower limit is
\begin{align}
    \frac{2\pi}{R}
    \sim
    10^{-16}\,{\rm MeV},
\end{align}
while the upper limit is estimated as
\begin{align}
    \frac{V_{\rm CV}}{D_{\nu}} \sim \frac{3\Omega}{\bar{\mu}_{\nu}l_{\rm mfp}}\sim 10^{-33}~{\rm MeV}\left(\frac{\Omega/2\pi}{1~{\rm Hz}}\right),
\end{align}
where we have substituted $l_{\rm mfp}\sim1\,{\rm cm}$~\cite{Kotake:2005zn,Yamamoto:2015gzz}.
Since the upper limit is much smaller than the lower limit, the CV-modes cannot propagate in the core of supernovae.

One main reason why the CV-mode cannot appear in supernovae while the CM-mode can is that the energy scale of the rotation ($\sim10^{-21}\,{\rm MeV}$) is much smaller than the possible scale of magnetic fields in neutron stars and supernovae ($\sim10^2 \,{\rm MeV}$).

\section{Angular dependence of the gravitational radiation due to the CM-mode and CV-mode}
\label{sec:angular}

As we mentioned, the energy density fluctuation $\delta\varepsilon$ is proportional to the number density fluctuation $\delta n$.
Since the CMW and CVW run along a specific direction such as the axis of the magnetic field or the rotational axis, gravitational waves from these sources are emitted in a certain direction.

To make our paper self-contained, we first summarize the quadrupole formula briefly. 
We introduce $\tilde{h}_{\alpha\beta} \equiv h_{\alpha\beta}-\frac{1}{2}h\eta_{\alpha\beta}$ satisfying the gauge fixing $\del_{\alpha}\tilde{h}^{\alpha\beta}=0$, where $h \equiv h^{\alpha}_{~\alpha}$ and $\eta_{\alpha\beta}={\rm diag}(-1,+1,+1,+1)$.
In the Minkowski space, the equation of motion for $\tilde{h}_{\alpha\beta}$ is
\begin{align}
    \Box \tilde{h}_{\alpha\beta}
    =
    -16\pi GT_{\alpha\beta},
\end{align}
where $\Box$ is the d'Alembertian and $T_{\alpha\beta}$ is the energy-momentum tensor.
The retarded solution can be written as
\begin{align}
    \tilde{h}_{\alpha\beta}(t,\bm{x})
    =
    4G\int \dif^{3}\bm{x}' \frac{T_{\alpha\beta}(t-|\bm{x}-\bm{x}'|,\bm{x}')}{|\bm{x}-\bm{x}'|}\,.
\end{align}
If we observe the gravitational wave at the distance $r=|\bm{x}-\bm{x'}|\gg R$ from the moving source whose characteristic size is $R$, we can rewrite the amplitude of the gravitational wave as
\begin{align}
\label{eq:amplitude-quadrupole}
    &\tilde{h}_{ab}
    =
    \frac{2G}{r}\frac{\del^{2}}{\del t^{2}}I_{ab}(t-r)\,,
    \\
    &I_{ab}
    =
    \int \dif^{3}\bm{x}' T^{00}(t,\bm{x}')x'_{a}x'_{b}\,,
\end{align}
where $I_{ab}$ is the quadrupole moment.

Let us consider the energy density fluctuation propagating in a certain direction. Here we assume that the fluctuation propagates along the $z$ axis with the constant speed $V_{\chi}$, where $V_{\chi} = V_{\rm CM}$ for the CM-mode and $V_{\chi} =  V_{\rm CV}$ for the CV-mode. In this case, its position $X_{a}(t)$ is given as
\begin{align}
    X_{a}(t)
    =
    (0,0,V_{\chi}t).
\end{align}
We can then write the fluctuation of the energy-momentum tensor,
\begin{align}
    \delta T^{00}(t,\bm{x})
    =
    \delta\varepsilon v\delta^{(3)}(\bm{x}-\bm{X}(t)),
\end{align}
where $\delta\varepsilon$ denotes the amplitude of the energy density fluctuation, and $v$ is a small volume element.
The quadrupole moment is
\begin{align}
    I_{ab}
    &=
    \int \dif^{3}\bm{x} \delta\varepsilon v\delta^{(3)}(\bm{x}-\bm{X}(t))x_{a}x_{b}
    \nonumber \\
    &=
    \delta\varepsilon vX_{a}X_{b}.
\end{align}
Since $X_{a}$ has only the $z$-component, all the components of the quadrupole moment except for $I_{zz}$ vanish, and 
\begin{align}
    I_{zz}
    =
    \delta\varepsilon v V_{\chi}^{2}t^{2}.
\end{align}
Substituting this into eq.~(\ref{eq:amplitude-quadrupole}), the amplitude of the gravitational wave reduces to
\begin{align}
    \tilde{h}_{zz}
    =
    \frac{4G \delta\varepsilon v}{r}V_{\chi}^{2}\,.
\end{align}
Using the correspondence between the Cartesian and polar coordinates,
$x = r\sin{\theta}\cos{\phi}$, $y = r\sin{\theta}\sin{\phi}$, $z = r\cos{\theta}$, we can write the amplitude as
\begin{align}
    \tilde{h}_{\theta\theta}
    =
    4Gvr \delta\varepsilon V_{\chi}^{2}\sin^{2}{\theta}.
\end{align}
From this $\theta$-dependence, the radiation of gravitational waves of the CM-mode and CV-mode becomes maximum at $\theta=\pi/2$.

\section{Discussions}
\label{sec:discussions}

In this paper, we have studied new kinds of seismic oscillations and resulting gravitational waves in neutron stars and supernovae, the CM-mode and CV-mode due to the chirality of quarks and neutrinos, respectively. 
Though the CM-mode in electron matter is strongly damped due to the electric conductivity (see appendix~\ref{app:CMW_electon_matter}), the CM-mode can exist in (two-flavor) quark matter.
The CV-mode in the core of supernovae is diffusive because the speed of the CVW, which is proportional to the angular frequency of the star, is too small. 
Our main results of the dispersion relations of the CM-mode and CV-mode are given in eqs.~(\ref{eq:dispersion-CM-expansion}) and (\ref{eq:dispersion-CV}). 
We also estimate the frequency and amplitude of the gravitational wave of the CM-mode for the 2SC phase showed in eqs.~(\ref{eq:CM_freq_quark}) and (\ref{eq:CM_amp_quark}).

Since the CM-mode has the explicit dependence on the magnetic field in neutron stars (see eq.~(\ref{eq:frequency-CM-parameter})) and can exist only in quark matter with nearly gapless quarks, possible detection of its signals in future observations would provide information about not only the magnetic field but also quark matter inside these compact stars. 
The gravitational waves of the CM-mode could be observed by a future detector with high sensitivity around the frequencies much larger than $10^4$~Hz.
As seismic oscillations cause electromagnetic emission, the CM-mode might also be observed as electromagnetic waves.%
\footnote{A similar example is the X-ray quasi-periodic oscillation, which is considered to originate from the seismic oscillations of the giant flare~\cite{Israel:2005av,Strohmayer:2006py}.}

There are several future directions. 
First, in this paper, we limit ourselves to the fluctuations of number density, chiral charge density, and energy density for simplicity. 
More generically, one should consider the fluctuation of the momentum density.  
Second, one should explore the effects of the magnetic field on the chirality flipping rate and diffusion coefficient that we have ignored here. Qualitatively, we expect that as spins of quarks tend to align along the magnetic field, the magnetic field suppresses the chirality flipping, and consequently, the lower limit of the frequency of the CM-mode decreases.
Third, it would be important to compute the frequencies of the CM-mode and CV-mode numerically by solving the Einstein equations for the metric without the Cowling approximation.

\acknowledgments
We thank Hajime Sotani for useful conversations and Igor Shovkovy for a critical comment on the earlier version of the manuscript.
This work was supported by the Keio Institute of Pure and Applied Sciences (KiPAS) project at Keio University, JSPS KAKENHI Grant Number~JP19K03852, and JST SPRING, Grant Number~JPMJSP2123.

\appendix

\section{CMW in electron matter}
\label{app:CMW_electon_matter}
In this appendix, we show that the CMW in electron matter is strongly damped by the electric conductivity $\sigma$~\cite{Rybalka:2018uzh,Shovkovy:2018tks}.

Unlike the case of the quark number current in section~\ref{sec:CM-mode_q}, the electric current in electron matter obtains an additional contribution of the Ohmic current induced by electric fields, $-e\bm{j}_{\rm e}=\sigma\bm{E}$. The fluctuations of the electron number and axial currents are expressed as
\begin{align}
    &\delta \bm{j}_{\rm e}
    =
    -\frac{e\bm{B}}{2\pi^2 \chi_{\rm e}}\delta n_{\rm 5,e}
    -D_{\rm e}\bm{\nabla}\delta n_{\rm e}
    -\frac{\sigma}{e}\delta\bm{E}\,,
    \\
    &\delta \bm{j}_{\rm 5,e}
    =
    -\frac{e\bm{B}}{2\pi^2 \chi_{\rm e}}\delta n_{\rm e}
    -D_{\rm e}\bm{\nabla}\delta n_{\rm 5,e}\,.
\end{align}
The electric field can be expressed as $E^{\hat{a}}=F^{\hat{t}\hat{a}}$, where $F^{\hat{t}\hat{a}}\equiv \del^{\hat{t}}A^{\hat{a}}-\del^{\hat{a}}A^{\hat{t}}$ is the field strength with $A^{\alpha}$ being the gauge field.
We use the same notations like the electron number density $n_{\rm e}$ as the main text except for the index $\rm e$ for electrons.
Since we have to treat the electric field as the dynamical field, we also use the Gauss law:
\begin{align}
    \nabla_{\alpha}\delta F^{t\alpha}=-e\delta j^{t}\,.
\end{align}
In the same way as section~\ref{sec:CM-mode_q}, we focus only on the direction of the magnetic field $\bm{B}=B\bm{e}_{\hat{z}}$.
We also assume that the electric field has only $z$-component, and then the gauge field is given as $\delta A^{\alpha}=(\e^{-\lambda}\delta A^{0},0,0,\e^{-\rho}\delta A^{3})$.
In this case, we can rewrite the Gauss law as
\begin{align}
    \del_{z}\delta E
    =
    -e\e^{\rho}\delta n_{\rm e}\,.
\end{align}

Including the effect of the chiral anomaly, the continuity equations can be expressed as
\begin{align}
    &\del_{t}\delta j_{\rm e}^t+\del_{z}\delta j_{\rm e}^{z}=0\,,
    \\
    &\del_{t}\delta j_{\rm 5,e}^t+\del_{z}\delta j_{\rm 5,e}^{z}=-\Gamma_{\rm flip}\delta j_{\rm 5,e}^{t}+\frac{e^2}{2\pi^2}B\delta E\,.
\end{align}
Eliminating $\delta E$ by the Gauss law, we obtain the following equations:
\begin{align}
    &\left(\omega+\im\e^{\lambda-\rho}D_{\rm e} k_{z}^2+\im\e^{\lambda}\sigma\right)\delta n_{\rm e}
    +\e^{\lambda-\rho}\frac{eB}{2\pi^2 \chi_{\rm e}}k_{z}\delta n_{\rm 5,e}
    =
    0\,,
    \\
    &\left(\e^{\lambda-\rho}\frac{eB}{2\pi^2 \chi_{\rm e}}k_{z}+\e^{\rho}\frac{e^3B}{2\pi^2 k_{z}}\right)\delta n_{\rm e}
    +\left(\omega+\im\e^{\lambda-\rho}D_{\rm e} k_{z}^2+\im\Gamma_{\rm flip}\right)\delta n_{\rm 5,e}
    =
    0\,.
\end{align}
The dispersion relation of the CM-mode of electron matter is written as
\begin{align}
    \omega_{\rm CM}
    &=
    \sqrt{V_{\rm CM}^2\left(k_{z}^2+\e^{\rho}e^2 \chi_{\rm e}\right)-\left(\e^{\lambda}\frac{\sigma}{2}-\frac{\Gamma_{\rm flip}}{2}\right)^2}
    \nonumber \\
    &\quad
    -\im\e^{\lambda}\frac{\sigma}{2}
    -\im\frac{\Gamma_{\rm flip}}{2}
    -\im\e^{\lambda-\rho}D_{\rm e}k_{z}^2\,,
\end{align}
where we have defined $V_{\rm CM}\equiv eB/(2\pi^2 \chi_{\rm e})$.
The CM-mode in electron matter has the gap because of the chiral anomaly.
This expression shows that the following inequality has to be satisfied for the propagation of the CM-mode:
\begin{align}
\label{eq:overdamped_condition}
    V_{\rm CM}^2\left(k_{z}^2+\e^{\rho}e^2 \chi_{\rm e}\right)
    >\left(\e^{\lambda}\frac{\sigma}{2}-\frac{\Gamma_{\rm flip}}{2}\right)^2\,.
\end{align}
To check whether this is satisfied, we estimate the order of magnitude of $\sigma$ and $\Gamma_{\rm flip}$.

We focus on the contribution of the Rutherford scattering (electron-proton scattering) to estimate the transport coefficients.
The other possible scattering processes are the Compton scattering and electron-electron scattering. 
However, the Compton scattering is not so efficient compared with the Rutherford scattering because the photon density $(\sim T^3)$ is much smaller than the proton density $(\sim\mu_{\rm e}^3)$.
Also, the electron-electron scattering has a minor contribution since the electrons are more degenerate than the protons in supernovae; even in neutron stars where protons are degenerate, the electron-electron scattering is parametrically suppressed compared with the Rutherford scattering because the typical energy of the electron ($\sim 10^2 ~{\rm MeV}$) is smaller than the proton mass ($\sim 10^{3}~{\rm MeV}$). 
Therefore, the Rutherford scattering is dominant.
The proton recoil can be ignored because the typical electron energy is much smaller than the proton mass $(\mu_{\rm e}\ll m_{\rm p})$.

\begin{figure}[ht]
\centering
\includegraphics[width=8cm]{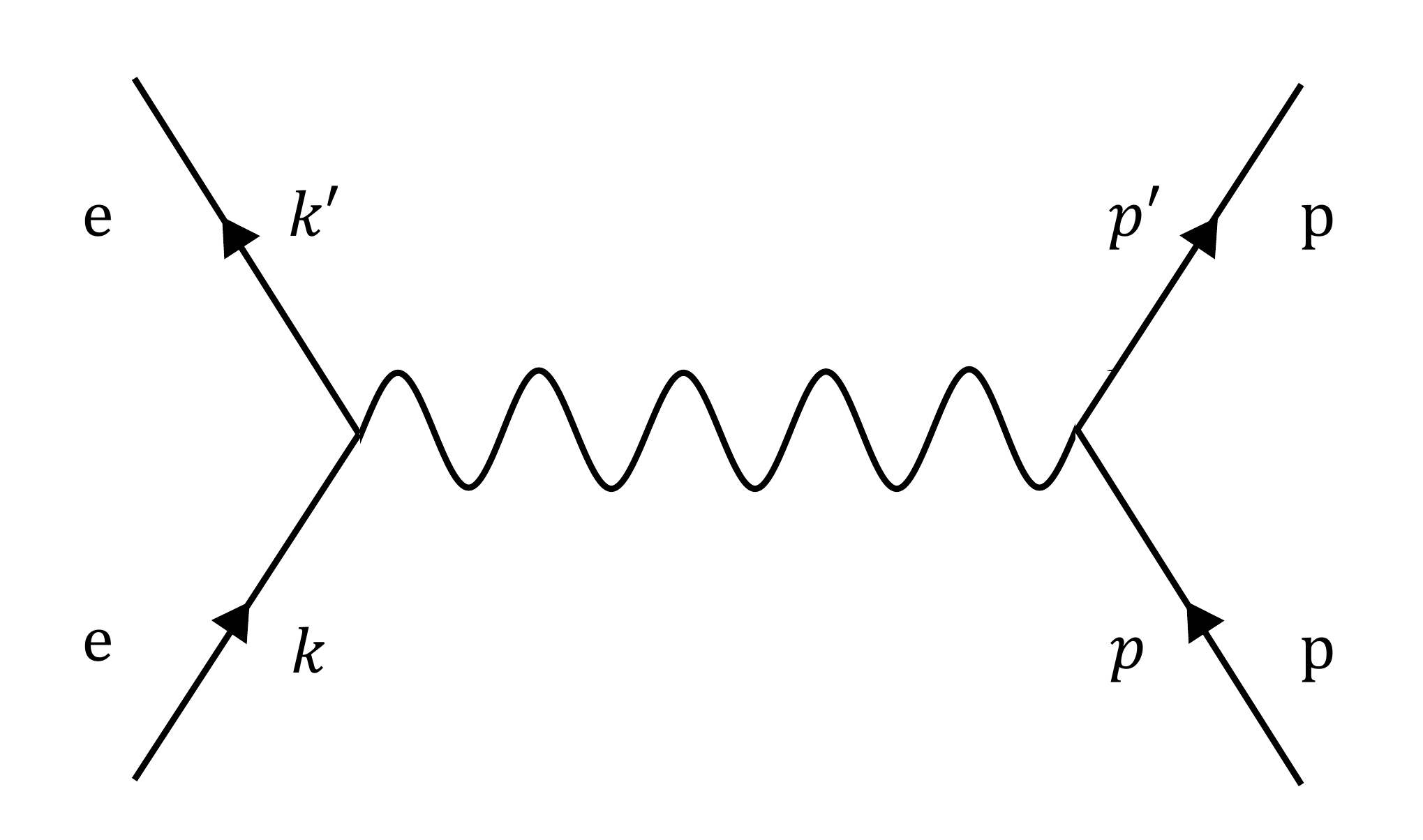}
\caption{The Feynman diagram of the Rutherford scattering.
In this figure, the incoming and outgoing electrons (protons) have the momenta $k$ and $k'$ ($p$ and $p'$), respectively.}
\label{fig:e-p-diagram}
\end{figure}

In the case of neutron stars $(T\sim 10^{6}\text{--}10^{9}~{\rm K})$, both electrons and protons are degenerate, and in the case of supernovae ($T\sim 10^{10}\text{--}10^{11}~{\rm K}$), electrons are degenerate while protons are non-degenerate and obey the Boltzmann distribution.

The electric conductivity is given by
\begin{align}
\label{eq:sigma}
    \sigma
    \sim
    \frac{e^2 n_{\rm e} l_{\rm mfp}}{\bar{\mu}_{\rm e}}\,.
\end{align}
We evaluate $l_{\rm mfp}$ under the isoenegetic approximation using the definition of the mean free path in ref.~\cite{Reddy:1997yr}, $1/l_{\rm mfp} \equiv \sigma_{\rm cross}/\mathcal{V}$, where $\sigma_{\rm cross}$ is the cross section of the Rutherford scattering (figure~\ref{fig:e-p-diagram}) and $\mathcal{V}$ is the volume of the system:
\begin{align}
\label{eq:mfp_e}
    \frac{1}{l_{\rm mfp}}
    &=
    2\int\frac{\dif^3 \bm{k}'}{(2\pi)^3}\frac{\dif^3 \bm{p}}{(2\pi)^3}\frac{\dif^3 \bm{p}'}{(2\pi)^3}
    (2\pi)^4 \delta^{(4)}(k^{\alpha}+p^{\alpha}-k'^{\alpha}-p'^{\alpha})|M|^2
    \nonumber \\
    &\qquad\times
    f_{\rm p}(p)[1-f_{\rm e}(k')][1-f_{\rm p}(p')]
    \nonumber \\
    &\simeq
    \left\{
    \begin{array}{ll}
    \displaystyle
    \frac{\alpha_{\rm e} m_{\rm p} T}{2\bar{\mu}_{\rm e}}
    \sim 
    10^{-5}~{\rm MeV}\left(\frac{T}{10^{6}~{\rm K}}\right)
    & \quad 
    (\text{neutron stars})
    \vspace{1em}
    \\
    \displaystyle
    \frac{\alpha_{\rm e} \bar{\mu}_{\rm e}}{6}
    \sim 
    10^{-1}~{\rm MeV}
    & \quad 
    (\text{supernovae})~~~~,
    \end{array}
    \right.
\end{align}
where we set $\bar{\mu}_{\rm e}\sim 10^{2}~{\rm MeV}$. When protons are non-degenerate in supernovae, the Pauli blocking of protons does not work, and the temperature dependence changes from $\sim T^1$ to $\sim T^0$.
Using eq.~(\ref{eq:sigma}), we can evaluate $\sigma$ as
\begin{align}
    \sigma
    &\sim
    \left\{
    \begin{array}{ll}
    \displaystyle
    10^{9}~{\rm MeV}\left(\frac{T}{10^{6}~{\rm K}}\right)^{-1}
    & \quad 
    (\text{neutron stars})
    \vspace{1em}
    \\
    \displaystyle
    10^{5}~{\rm MeV}
    & \quad 
    (\text{supernovae})~~~~.
    \end{array}
    \right.
\end{align}
The chirality flipping rate in electron matter~\cite{Dvornikov:2015iua,Grabowska:2014efa}%
\footnote{In the case of neutron stars, our expression differs from the result in~ref.~\cite{Dvornikov:2015iua} where the plasma frequency instead of the Debye momentum is used for the IR cutoff. However, as the effective mass of the photon originating from the longitudinal part of the photon self-energy is the Debye momentum in the regime where the proton recoil is ignored, we here use the Debye momentum for the IR cutoff.} is given by
\begin{align}
    \Gamma_{\rm flip}\simeq
    \left\{
    \begin{array}{ll}
    \displaystyle
    \frac{\alpha^2 m_{\rm e}^2 m_{\rm p} T}{\pi \bar{\mu}_{\rm e}^3}\left(\ln{\frac{4\bar{\mu}_{\rm e}^2}{q_{\rm s}^2}}-1\right)
    \sim 
    10^{-12}~{\rm MeV}\left(\frac{T}{10^{6}~{\rm K}}\right)
    & \quad 
    (\text{neutron stars})
    \vspace{1em}
    \\
    \displaystyle
    \frac{\alpha^2 m_{\rm e}^2}{3\pi \bar{\mu}_{\rm e}}\left(\ln{\frac{4\bar{\mu}_{\rm e}^2}{q_{\rm s}^2}}-1\right)
    \sim 
    10^{-8}~{\rm MeV}
    & \quad 
    (\text{supernovae})~~~~.
    \end{array}
    \right.
\end{align}

As $k_{z}\ll 2\pi/l_{\rm mfp}\ll \bar{\mu}_{\rm e}$ and $\sigma\gg \Gamma_{\rm flip}$, the condition (\ref{eq:overdamped_condition}) can be expressed parametrically as
\begin{align}
    1
    \gtrsim
    \frac{\bar{\mu}_{\rm e}\sigma}{\alpha_{\rm e} B}
    \sim
    \left\{
    \begin{array}{ll}
    \displaystyle
    10^{9}~\left(\frac{B}{10^{18}~{\rm Gauss}}\right)^{-1}\left(\frac{T}{10^{6}~{\rm K}}\right)^{-1}
    & \quad 
    (\text{neutron stars})
    \vspace{1em}
    \\
    \displaystyle
    10^{5}~\left(\frac{B}{10^{18}~{\rm Gauss}}\right)^{-1}
    & \quad 
    (\text{supernovae})~~~~.
    \end{array}
    \right.
\end{align}
To satisfy the inequality, the magnetic field has to be at least larger than $10^{23}~{\rm Gauss}$.
On the other hand, the magnetic field inside neutron stars and supernovae is at most about $10^{18}\text{--}10^{19}~{\rm Gauss}$ due to the virial theorem~\cite{Lai:1991}.
Therefore, the CM-mode in electron matter is strongly damped. 

\section{Helicity eigenspinor}
\label{app:helicity_spinor}

In this appendix, we derive the helicity eigenspinors $u_{h}(k)$ of fermions which are used to obtain the currents in section~\ref{sec:CM-mode_q}.
The eigenspinors satisfy the following equations,
\begin{align}
\label{eq:App:scat-amp:Dirac-eq}
    &(\slashed{k}-m_{\rm q})u_{h}(k)
    =
    0,
    \\
\label{eq:App:scat-amp:helicity-eq}
    &\left(\frac{\bm{\Sigma}\cdot\bm{k}}{|\bm{k}|}-h\right)u_{h}(k)
    =
    0\,,
\end{align}
with the normalization $\bar u_{h} u_{h'} = 2m_{\rm q} \delta_{h h'}$,
where $m_{\rm q}$ is the quark mass, $h=\pm1$ is the helicity, and $\Sigma^a$ is defined as
\begin{align}
    \Sigma^a
    =
    \left(
    \begin{array}{cc}
        \sigma^{a} & 0 \\
        0 & \sigma^{a}
    \end{array}
    \right),
\end{align}
with $\sigma^{a}$ ($a=1,2,3$) being the Pauli matrices.
Using the Dirac representation of the gamma matrices,
\begin{align}
    \gamma^{0}
    =
    \left(
    \begin{array}{cc}
        1 & 0 \\
        0 & -1
    \end{array}
    \right),
    \qquad
    \gamma^{a}
    =
    \left(
    \begin{array}{cc}
        0 & \sigma^{a} \\
        -\sigma^{a} & 0
    \end{array}
    \right)\,,
\end{align}
eq.~(\ref{eq:App:scat-amp:Dirac-eq}) can be written as
\begin{align}
    \left(
    \begin{array}{cc}
        k-m_{\rm q} & -\bm{\sigma}\cdot\bm{k} \\
        \bm{\sigma}\cdot\bm{k} & -k-m_{\rm q}
    \end{array}
    \right)
    u_{h}(k)
    =
    0\,.
\end{align}
The solution of this equation is given by
\begin{align}
    u_{h}(k)
    =
    N(k)
    \left(
    \begin{array}{c}
        \xi_{h} \\
        hF(k)\xi_{h}
    \end{array}
    \right)\,,
\end{align}
where $\xi_{h}$ is a two-component spinor satisfying 
\begin{align}
    \left(\bm{\sigma}\cdot\bm{k}-hk\right)\xi_{h}
    =
    0\,,
\end{align}
and 
\begin{align}
    N(k)
    \equiv
    \sqrt{\epsilon_k+m_{\rm q}}\,,
    \qquad
    F(k)
    \equiv
    \sqrt{\frac{\epsilon_k-m_{\rm q}}{\epsilon_k+m_{\rm q}}}\,,
\end{align}
with $\epsilon_k = \sqrt{k^2 + m_{\rm q}^2}$.
In the spherical coordinates, the solution of this equation is given by
\begin{align}
    &\xi_{+}
    =
    \left(
    \begin{array}{c}
        \e^{-\im\phi/2}\cos{\frac{\theta}{2}}
        \vspace{0.5em}
        \\
        \e^{\im\phi/2}\sin{\frac{\theta}{2}} 
    \end{array}
    \right)\,,
    \\
    &\xi_{-}
    =
    \left(
    \begin{array}{c}
        -\e^{-\im\phi/2}\sin{\frac{\theta}{2}}
        \vspace{0.5em}
        \\
        \e^{\im\phi/2}\cos{\frac{\theta}{2}} 
    \end{array}
    \right)\,.
\end{align}



\begin{thebibliography}{15}

\bibitem{LIGOScientific:2016aoc}
B.~P.~Abbott \textit{et al.} [LIGO Scientific and Virgo],
``Observation of Gravitational Waves from a Binary Black Hole Merger,''
Phys. Rev. Lett. \textbf{116}, 061102 (2016)
[arXiv:1602.03837 [gr-qc]].

\bibitem{LIGOScientific:2017vwq}
B.~P.~Abbott \textit{et al.} [LIGO Scientific and Virgo],
``GW170817: Observation of Gravitational Waves from a Binary Neutron Star Inspiral,''
Phys. Rev. Lett. \textbf{119}, 161101 (2017)
[arXiv:1710.05832 [gr-qc]].

\bibitem{Vilenkin:1980fu}
A.~Vilenkin,
``Equilibrium parity-violating current in a magnetic field,''
Phys. Rev. D \textbf{22}, 3080 (1980).

\bibitem{Nielsen:1983rb}
H.~B.~Nielsen and M.~Ninomiya,
``The Adler-Bell-Jackiw anomaly and Weyl fermions in a crystal,''
Phys. Lett. B \textbf{130}, 389 (1983).

\bibitem{Fukushima:2008xe}
K.~Fukushima, D.~E.~Kharzeev and H.~J.~Warringa,
``Chiral magnetic effect,''
Phys. Rev. D \textbf{78}, 074033 (2008)
[arXiv:0808.3382 [hep-ph]].

\bibitem{Vilenkin:1979ui}
A.~Vilenkin,
``Macroscopic parity-violating effects: Neutrino fluxes from rotating black holes and in rotating thermal radiation,''
Phys. Rev. D \textbf{20}, 1807 (1979).

\bibitem{Son:2009tf}
D.~T.~Son and P.~Sur\'owka,
``Hydrodynamics with Triangle Anomalies,''
Phys. Rev. Lett. \textbf{103}, 191601 (2009)
[arXiv:0906.5044 [hep-th]].

\bibitem{Landsteiner:2011cp}
K.~Landsteiner, E.~Megias and F.~Pena-Benitez,
``Gravitational Anomaly and Transport,''
Phys. Rev. Lett. \textbf{107}, 021601 (2011)
[arXiv:1103.5006 [hep-ph]].

\bibitem{Newman:2005hd}
G.~M.~Newman,
``Anomalous hydrodynamics,''
JHEP \textbf{01}, 158 (2006)
[arXiv:hep-ph/0511236 [hep-ph]].

\bibitem{Kharzeev:2010gd}
D.~E.~Kharzeev and H.~U.~Yee,
``Chiral magnetic wave,''
Phys. Rev. D \textbf{83}, 085007 (2011)
[arXiv:1012.6026 [hep-th]].

\bibitem{Jiang:2015cva}
Y.~Jiang, X.~G.~Huang and J.~Liao,
``Chiral vortical wave and induced flavor charge transport in a rotating quark-gluon plasma,''
Phys. Rev. D \textbf{92}, 071501 (2015)
[arXiv:1504.03201 [hep-ph]].

\bibitem{Yamamoto:2015ria}
N.~Yamamoto,
``Chiral Alfv\'en Wave in Anomalous Hydrodynamics,''
Phys. Rev. Lett. \textbf{115}, 141601 (2015)
[arXiv:1505.05444 [hep-th]].

\bibitem{Chernodub:2015gxa}
M.~N.~Chernodub,
``Chiral heat wave and mixing of magnetic, vortical and heat waves in chiral media,''
JHEP \textbf{01}, 100 (2016)
[arXiv:1509.01245 [hep-th]].

\bibitem{Sen:2016jzl}
S.~Sen and N.~Yamamoto,
``Chiral Shock Waves,''
Phys. Rev. Lett. \textbf{118}, 181601 (2017)
[arXiv:1609.07030 [hep-th]].

\bibitem{Kharzeev:2015znc}
D.~E.~Kharzeev, J.~Liao, S.~A.~Voloshin and G.~Wang,
``Chiral magnetic and vortical effects in high-energy nuclear collisions\textemdash{}A status report,''
Prog. Part. Nucl. Phys. \textbf{88}, 1 (2016)
[arXiv:1511.04050 [hep-ph]].

\bibitem{Rybalka:2018uzh}
D.~O.~Rybalka, E.~V.~Gorbar and I.~A.~Shovkovy,
``Hydrodynamic modes in a magnetized chiral plasma with vorticity,''
Phys. Rev. D \textbf{99}, 016017 (2019)
[arXiv:1807.07608 [hep-th]].

\bibitem{Shovkovy:2018tks}
I.~A.~Shovkovy, D.~O.~Rybalka and E.~V.~Gorbar,
PoS \textbf{Confinement2018}, 029 (2018)
[arXiv:1811.10635 [nucl-th]].

\bibitem{Grabowska:2014efa}
D.~Grabowska, D.~B.~Kaplan and S.~Reddy,
``Role of the electron mass in damping chiral plasma instability in Supernovae and neutron stars,''
Phys. Rev. D \textbf{91}, 085035 (2015)
[arXiv:1409.3602 [hep-ph]].

\bibitem{Dvornikov:2015iua}
M.~Dvornikov,
``Relaxation of the chiral imbalance and the generation of magnetic fields in magnetars,''
J. Exp. Theor. Phys. \textbf{123}, 967 (2016)
[arXiv:1510.06228 [hep-ph]].

\bibitem{Boyarsky:2020cyk}
A.~Boyarsky, V.~Cheianov, O.~Ruchayskiy and O.~Sobol,
``Evolution of the Primordial Axial Charge across Cosmic Times,''
Phys. Rev. Lett. \textbf{126}, 021801 (2021)
[arXiv:2007.13691 [hep-ph]].

\bibitem{Boyarsky:2020ani}
A.~Boyarsky, V.~Cheianov, O.~Ruchayskiy and O.~Sobol,
``Equilibration of the chiral asymmetry due to finite electron mass in electron-positron plasma,''
Phys. Rev. D \textbf{103}, 013003 (2021)
arXiv:2008.00360 [hep-ph]].

\bibitem{Cowling:1941}
T.~G.~Cowling,
``The non-radial oscillations of polytropic stars,"
Mon.\ Not.\ Roy.\ Astron.\ Soc.\ {\bf 101},(1941).

\bibitem{Papaloizou:1978zz}
J.~Papaloizou and J.~E.~Pringle,
``Non-radial oscillations of rotating stars and their relevance to the short-period oscillations of cataclysmic variables,''
Mon. Not. Roy. Astron. Soc. \textbf{182}, 423-442 (1978).

\bibitem{Saio:1982}
H.~Saio,
``R-mode oscillation in uniformly rotating stars,''
Astro. Phys. J. \textbf{256}, 717, (1982).

\bibitem{McDermott:1983}
P.~N.~McDermott, H.~M.~Van~Horn and J.~f.~Scholl,
``Nonradial g-mode oscillations of warm neutron stars,"
Astro.\ Phys.\ {\bf 268},(1983).

\bibitem{Andersson:1997rn}
N.~Andersson and K.~D.~Kokkotas,
``Towards gravitational wave asteroseismology,''
Mon. Not. Roy. Astron. Soc. \textbf{299}, 1059 (1998)
[arXiv:gr-qc/9711088 [gr-qc]].

\bibitem{Kokkotas:1991}
K.~D.~Kokkotas and B.~F.~Schutz,
``W-modes : A new family of normal modes of pulsating relativistic stars,''
Mon. Not. R. Astron. Soc. \textbf{255}, 119, (1991).

\bibitem{Andersson:1996}
N.~Andersson, K.~D.~Kokkotas ans B.~F.~Schutz,
``Space-time modes of relativistic stars,"
Mon. Not. R. Astron. Soc. \textbf{280}, 1230, (1996).

\bibitem{Kokkotas:1986}
K.~D.~Kokkotas and B.~F.~Schutz,
``Normal modes of a model radiating system,''
Mon. Not. R. Astron. Soc. \textbf{18}, 913, (1986)

\bibitem{Kokkotas:1999}
K.~D.~Kokkotas and B.~G.~Schmidt,
``Quasi-normal modes of stars and black holes,"
Living Rev.~Relativ. \textbf{2}, (1999).

\bibitem{Andersson:2000mf}
N.~Andersson and K.~D.~Kokkotas,
``The R mode instability in rotating neutron stars,''
Int. J. Mod. Phys. D \textbf{10}, 381 (2001)
[arXiv:gr-qc/0010102 [gr-qc]].

\bibitem{Yoshida:1997bf}
S.~Yoshida and Y.~Kojima,
``Accuracy of the relativistic Cowling approximation in slowly rotating stars,''
Mon. Not. Roy. Astron. Soc. \textbf{289}, 117 (1997)
[arXiv:gr-qc/9705081 [gr-qc]].

\bibitem{Sotani:2020mwc}
H.~Sotani and T.~Takiwaki,
``Accuracy of relativistic Cowling approximation in protoneutron star asteroseismology,''
Phys. Rev. D \textbf{102}, 063025 (2020)
[arXiv:2009.05206 [astro-ph.HE]].

\bibitem{Son:2004tq}
D.~T.~Son and A.~R.~Zhitnitsky,
``Quantum anomalies in dense matter,''
Phys. Rev. D \textbf{70}, 074018 (2004)
[arXiv:hep-ph/0405216 [hep-ph]].

\bibitem{Metlitski:2005pr}
M.~A.~Metlitski and A.~R.~Zhitnitsky,
``Anomalous axion interactions and topological currents in dense matter,''
Phys. Rev. D \textbf{72}, 045011 (2005)
[arXiv:hep-ph/0505072 [hep-ph]].

\bibitem{Kharzeev:2010gr}
D.~E.~Kharzeev and D.~T.~Son,
``Testing the chiral magnetic and chiral vortical effects in heavy ion collisions,''
Phys. Rev. Lett. \textbf{106}, 062301 (2011)
[arXiv:1010.0038 [hep-ph]].

\bibitem{Barrois:1977xd}
B.~C.~Barrois,
``Superconducting Quark Matter,''
Nucl. Phys. B \textbf{129}, 390-396 (1977).

\bibitem{Bailin:1983bm}
D.~Bailin and A.~Love,
``Superfluidity and Superconductivity in Relativistic Fermion Systems,''
Phys. Rept. \textbf{107}, 325 (1984).

\bibitem{Alford:1997zt}
M.~G.~Alford, K.~Rajagopal and F.~Wilczek,
``QCD at finite baryon density: Nucleon droplets and color superconductivity,''
Phys. Lett. B \textbf{422}, 247-256 (1998)
[arXiv:hep-ph/9711395 [hep-ph]].

\bibitem{Rapp:1997zu}
R.~Rapp, T.~Sch\"afer, E.~V.~Shuryak and M.~Velkovsky,
``Diquark Bose condensates in high density matter and instantons,''
Phys. Rev. Lett. \textbf{81}, 53-56 (1998)
[arXiv:hep-ph/9711396 [hep-ph]].

\bibitem{Adler:1969gk}
S.~L.~Adler,
``Axial vector vertex in spinor electrodynamics,''
Phys. Rev. \textbf{177}, 2426 (1969).

\bibitem{Bell:1969ts}
J.~S.~Bell and R.~Jackiw,
``A PCAC puzzle: $\pi^0 \to \gamma \gamma$ in the $\sigma$ model,''
Nuovo Cim. A \textbf{60}, 47 (1969).

\bibitem{Son:2012wh}
D.~T.~Son and N.~Yamamoto,
``Berry Curvature, Triangle Anomalies, and the Chiral Magnetic Effect in Fermi Liquids,''
Phys. Rev. Lett. \textbf{109}, 181602 (2012)
[arXiv:1203.2697 [cond-mat.mes-hall]].

\bibitem{Son:2012zy}
D.~T.~Son and N.~Yamamoto,
``Kinetic theory with Berry curvature from quantum field theories,''
Phys. Rev. D \textbf{87}, 085016 (2013)
[arXiv:1210.8158 [hep-th]].

\bibitem{Schafer:2002ty}
T.~Sch\"afer,
``Instanton effects in QCD at high baryon density,''
Phys. Rev. D \textbf{65}, 094033 (2002)
[arXiv:hep-ph/0201189 [hep-ph]].

\bibitem{Heiselberg:1993cr}
H.~Heiselberg and C.~J.~Pethick,
``Transport and relaxation in degenerate quark plasmas,''
Phys. Rev. D \textbf{48}, 2916 (1993).

\bibitem{Kogut:2004su}
J.~B.~Kogut and M.~A.~Stephanov,
``The phases of quantum chromodynamics: From confinement to extreme environments,''
Camb. Monogr. Part. Phys. Nucl. Phys. Cosmol. \textbf{21}, Cambridge University Press (2004).

\bibitem{Lihshitz:1980}
E.~M.~Lifshitz and L.~P.~Pitaevskii,
``Statistical Physics, Part 2,''
Pergamon Press (1980).

\bibitem{Schafer:2001za}
T.~Sch\"afer,
Phys. Rev. D \textbf{65}, 074006 (2002)
[arXiv:hep-ph/0109052 [hep-ph]].

\bibitem{Hong:1998tn}
D.~K.~Hong,
``An Effective field theory of QCD at high density,''
Phys. Lett. B \textbf{473}, 118-125 (2000)
[arXiv:hep-ph/9812510 [hep-ph]].

\bibitem{Hong:1999ru}
D.~K.~Hong,
``Aspects of high density effective theory in QCD,''
Nucl. Phys. B \textbf{582}, 451-476 (2000)
[arXiv:hep-ph/9905523 [hep-ph]].

\bibitem{Andersson:1996pn}
N.~Andersson and K.~D.~Kokkotas,
``Gravitational waves and pulsating stars: What can we learn from future observations?,''
Phys. Rev. Lett. \textbf{77}, 4134 (1996)
[arXiv:gr-qc/9610035 [gr-qc]].

\bibitem{Sato:1974}
K.~Sato,
``Neutrino Degeneracy in Supernova
Cores and Neutral Current
of Weak Interaction,''
Prog. Theor. Phys. \textbf{53} (1975).

\bibitem{Kotake:2005zn}
K.~Kotake, K.~Sato and K.~Takahashi,
``Explosion mechanism, neutrino burst, and gravitational wave in core-collapse supernovae,''
Rept. Prog. Phys. \textbf{69}, 971 (2006)
[arXiv:astro-ph/0509456 [astro-ph]].

\bibitem{Yamamoto:2015gzz}
N.~Yamamoto,
``Chiral transport of neutrinos in supernovae: Neutrino-induced fluid helicity and helical plasma instability,''
Phys. Rev. D \textbf{93}, 065017 (2016)
[arXiv:1511.00933 [astro-ph.HE]].

\bibitem{Flachi:2017vlp}
A.~Flachi and K.~Fukushima,
``Chiral vortical effect with finite rotation, temperature, and curvature,''
Phys. Rev. D \textbf{98}, 096011 (2018)
[arXiv:1702.04753 [hep-th]].

\bibitem{Israel:2005av}
G.~Israel, T.~Belloni, L.~Stella, Y.~Rephaeli, D.~Gruber, P.~G.~Casella, S.~Dall'Osso, N.~Rea, M.~Persic and R.~Rothschild,
``Discovery of rapid x-ray oscillations in the tail of the SGR 1806-20 hyperflare,''
Astrophys. J. Lett. \textbf{628}, L53 (2005)
[arXiv:astro-ph/0505255 [astro-ph]].

\bibitem{Strohmayer:2006py}
T.~E.~Strohmayer and A.~L.~Watts,
``The 2004 Hyperflare from SGR 1806-20: Further Evidence for Global Torsional Vibrations,''
Astrophys. J. \textbf{653}, 593 (2006)
[arXiv:astro-ph/0608463 [astro-ph]].

\bibitem{Reddy:1997yr}
S.~Reddy, M.~Prakash and J.~M.~Lattimer,
``Neutrino interactions in hot and dense matter,''
Phys. Rev. D \textbf{58}, 013009 (1998)
[arXiv:astro-ph/9710115 [astro-ph]].

\bibitem{Lai:1991}
D.~Lai and S.~L.~Shapiro,
``Cold equation of state in a strong magnetic field: Effect of inverse $\beta$-decay,''
Astro. Phys. J. \textbf{383}, 745, (1991).

\end{thebibliography}
\end{document}